\documentclass[fleqn,usenatbib]{mnras}

\usepackage{newtxtext,newtxmath}
\usepackage[T1]{fontenc}
\usepackage{graphicx}
\usepackage{anyfontsize}
\usepackage{tabularx}
\usepackage{supertabular}
\usepackage[math]{cellspace}

\usepackage{xspace}
\usepackage[dvipsnames]{xcolor}
\usepackage{fontawesome}
\usepackage{multirow}
\usepackage[authoryear]{natbib}

\usepackage{orcidlink}

\usepackage{hyperref}
\hypersetup{colorlinks=true, allcolors=Blue}

\newcommand{\lenspop}{\texttt{lenspop}\xspace}
\newcommand{\spectacularlenses}{17\xspace} % Guillaume's lenses
\newcommand{\halflenses}{99\xspace} % lenses that scored 6/12
\newcommand{\threequarterlenses}{42\xspace} % lenses that scored 8/12

\newcommand{\Ntotal}{59\xspace} % i.e. 17 + 42
\newcommand{\Ntotalweak}{116\xspace} % i.e. 17 + 99

\newcommand{\catname}{COWLS\xspace}
\newcommand{\catfullname}{COSMOS-Web Lens Survey\xspace}

\let\oldcite\cite
\renewcommand{\cite}[1]{\mbox{\oldcite{#1}}}

\newcommand{\nht}[1]{{\color{black} {#1}}}

\newcommand{\LUPM}{%
Laboratoire Univers et Particules de Montpellier, CNRS \& Université de Montpellier, Parvis Alexander Grothendieck, Montpellier, France 34090
}

\newcommand{\Newcastle}{%
School of Mathematics, Statistics and Physics, Newcastle University, Herschel Building, Newcastle-upon-Tyne, NE1 7RU, UK
}

\newcommand{\DurhamICC}{%
Department of Physics, Institute for Computational Cosmology, Durham University, South Road, Durham DH1 3LE, UK
}

\newcommand{\DurhamCEA}{%
Department of Physics, Centre for Extragalactic Astronomy, Durham University, South Road, Durham DH1 3LE, UK
}

\newcommand{\Northeastern}{%
Department of Physics, Northeastern University, 360 Huntington Ave, Boston, MA USA}

\newcommand{\Liege}{%
STAR Institute, Quartier Agora - All\'ee du six Ao\^ut, 19c B-4000 Li\`ege, Belgium}

\newcommand{\Aalto}{%
Department of Computer Science, Aalto University, PO Box 15400, Espoo, FI-00 076, Finland
}
\newcommand{\Helsinki}{%
Department of Physics, Faculty of Science, University of Helsinki, 00014-Helsinki, Finland}

\newcommand{\UTAustin}{%
Department of Astronomy, The University of Texas at Austin, Austin, TX, USA
}

\newcommand{\DAWN}{%
Cosmic Dawn Centre (DAWN), Denmark
}

\newcommand{\NBI}{%
Niels Bohr Institute, University of Copenhagen, Jagtvej 128, 2200 Copenhagen, Denmark}

\newcommand{\JPL}{%
Jet Propulsion Laboratory, California Institute of Technology, 4800, Oak Grove Drive, Pasadena, CA, USA}

\newcommand{\PMO}{Purple Mountain Observatory, Chinese Academy of Sciences, 10
Yuanhua Road, Nanjing 210023, China}

\newcommand{\IAP}{Institut d’Astrophysique de Paris, UMR 7095, CNRS, Sorbonne Universit\'e, 98 bis boulevard Arago, F-75014 Paris, France}

\newcommand{\LAM}{Laboratoire d'astrophysique de Marseille, Aix Marseille University, CNRS, CNES, Marseille, France}

\newcommand{\UCSB}{Department of Phyiscs University of California Santa Barbara, CA,93106, CA}

\newcommand{\Rochester}{Laboratory for Multiwavelength Astrophysics, School of Physics and Astronomy, Rochester Institute of Technology, 84 Lomb Memorial Drive, Rochester, NY 14623, USA}

\newcommand{\STScI}{Space Telescope Science Institute, 3700 San Martin Drive, Baltimore, MD 21218, USA}

\newcommand{\UCSC}{Department of Astronomy and Astrophysics, University of California, Santa Cruz, 1156 High Street, Santa Cruz, CA 95064 USA}

\newcommand{\Hawaii}{Department of Physics and Astronomy, University of Hawaii, Hilo, 200 W Kawili St, Hilo, HI 96720, USA}

\newcommand{\UCR}{Department of Physics and Astronomy, University of California, Riverside, 900 University Avenue, Riverside, CA 92521, USA}

\newcommand{\Caltech}{Caltech/IPAC, 1200 E. California Blvd., Pasadena, CA 91125, USA}

\newcommand{\ICG}{Institute of Cosmology and Gravitation, University of Portsmouth, Dennis Sciama Building, Burnaby Road, Portsmouth, PO1 3FX, UK}

\title[\catname III: forecasts versus data]{The \catfullname (\catname) III: forecasts versus data}

\author[N.~B.~Hogg et al.]{Natalie~B.~Hogg,\orcidlink{0000-0001-9346-4477}$^{1}$\thanks{E-mail: natalie.hogg@lupm.in2p3.fr}
James~W.~Nightingale,$^{2}$
Quihan~He,$^{3}$ 
Jacqueline McCleary,$^{4}$ 
Guillaume Mahler,$^{3, 5, 6}$
\newauthor
Aristeidis Amvrosiadis,$^{3}$
Ghassem Gozaliasl,$^{7,8}$
Edward Berman,$^{4}$
Richard J. Massey,$^{6}$
Diana~Scognamiglio,$^{9}$
\newauthor
Maximilien Franco,$^{10}$
Daizhong Liu,$^{11}$
Marko Shuntov,$^{12, 13}$
Louise Paquereau,$^{14}$
Olivier Ilbert,$^{15}$
\newauthor
Natalie Allen,$^{12, 13}$
Sune Toft,$^{12, 13}$
Hollis B. Akins,$^{10}$
Caitlin M. Casey,$^{10,16,12}$
Jeyhan S. Kartaltepe,$^{17}$
\newauthor
Anton M. Koekemoer,$^{18}$
Henry Joy McCracken,$^{14}$
Jason D. Rhodes,$^{9}$
Brant E. Robertson,$^{19}$
\newauthor
Nicole E. Drakos,$^{20}$
Andreas L. Faisst,$^{21}$
Hossein Hatamnia,$^{22}$
and Sophie L. Newman$^{23}$
\\
Affiliations can be found after the references.
}
\date{}

\pubyear{2025}

\begin{document}
\label{firstpage}
\pagerange{\pageref{firstpage}--\pageref{lastpage}}
\maketitle

% abstract is currently 259 words; limit is 250
\begin{abstract}
{We compare forecasts for the abundance and properties of strong gravitational lenses in the COSMOS-Web survey, a $0.54$~deg$^2$ survey of the COSMOS field using the NIRCam and MIRI instruments aboard \textit{JWST}, with the first catalogue of strong lens candidates identified in the observed NIRCam data, \catname. We modify the \lenspop package to produce a forecast for strong lensing in COSMOS-Web. We add a new mock galaxy catalogue to use as the source population, as well as the COSMOS-Web survey specifications, including the transmission data for the four NIRCam filters used. We forecast 107 strong lenses can be detected in COSMOS-Web across all bands, assuming complete subtraction of the lens galaxy light. The majority of the lenses are forecast to have small Einstein radii ($\theta_{\rm E} < 1$ arcsecond) and lie at redshifts between $0 < z <2$, whilst the source redshift distribution peaks at $z\sim 3$ and has a long tail extending up to $z \sim 11$, unambiguously showing that strong lensing in \textit{JWST} can probe the entirety of the epoch of reionisation. We compare our forecast with the distributions of Einstein radii, lens photometric redshifts, and lens and source magnitudes in the observed lenses, finding that whilst the forecast and observed Einstein radii distributions match, the redshifts and magnitudes do not. The observed lens redshift distribution peaks at a slightly lower redshift than the forecast one, whilst the lens magnitudes are systematically brighter in the observed data than in the forecast. 
% Assuming all lenses present in the COSMOS-Web field have been discovered, this implies that the \lenspop model must be reassessed and further modified to better capture the COWLS population. On the other hand, 
% Assuming the validity of the \lenspop forecast, this finding motivates further inspection of the \catname data to search for the as-yet undiscovered population of faint, high-redshift lenses.
}
\end{abstract}

\begin{keywords} 
\nht{cosmology: observational} -- gravitational lensing: strong -- methods: statistical
\end{keywords}

\section{Introduction}

Strong gravitational lensing -- the deflection of light by a massive object which produces multiple images of the source -- provides a wealth of information about our Universe. The magnification of distant sources has facilitated the study of high-redshift galaxies \citep{Hezaveh2013, Swinbank2015}, whilst forward modelling of the lensed images can provide insights into the mass distribution of the inner part of lens galaxies \citep{Shajib2021, Tan2024} and the potential presence of low-mass dark sub-haloes within the main lens or along the line of sight \citep{Vegetti2012, ORiordan:2022qds}.

A lensed transient source allows the time delay between separate images to be determined, which provides a measurement of $H_0$, the expansion rate of the Universe at $z=0$ \citep{Refsdal1964b, Suyu:2013kha, Wong:2019kwg}. Bayesian hierarchical modelling enables constraints to be placed on other cosmological parameters, such as the equation of state of dark energy, $w$ \citep{Hogg:2023khs, Li2024}. A further cosmological application of strong lensing is the proposed measurement of the weak lensing distortions to strong lensing images \citep{Birrer:2016xku, Birrer:2017sge, Fleury:2021tke, Hogg:2022ycw, Duboscq:2024asf, Hogg:2025wac}, which may provide an independent and complementary constraint on the large-scale distribution of dark matter in the Universe to that provided by traditional cosmic shear measurements from galaxy surveys, e.g. \cite{HSCDR1, DESDR1, Heymans:2020gsg}.

These scientific applications of strong lensing are naturally limited by the number of strong lenses available to analyse, which is currently around $1,000$ \citep{Sonnenfeld2020}, but the very large numbers of lenses forecast to be observed in ongoing or forthcoming wide area surveys such as \textit{Euclid} and the Large Synoptic Survey Telescope (LSST) at the Vera C. Rubin Observatory should yield huge developments in these fields of study \citep{Sonnenfeld2021b}. For example, \textit{Euclid} alone is expected to find $\mathcal{O}(10^5)$ strong lenses \citep{Collett:2015roa, Holloway:2023axl, Ferrami:2024obm}, bringing strong lensing into the era of precision astrophysics and cosmology, with a \nht{large} number of strong lenses already identified in the Early Release Observations \citep{Euclid:2024jyk,Euclid:2024juc,ORiordan2025, Nagam2025} \nht{and the Quick Data Release, or Q1, observations \citep{EuclidQ1A, EuclidQ1B, EuclidQ1C, EuclidQ1D, EuclidQ1E}.}

Understanding the strong lensing science return that can be expected from a given survey relies on an accurate forecast of the number of strong lenses that a survey will observe. However, accurately forecasting the abundance and properties of strong gravitational lenses can be difficult. Firstly, the number of strong lenses that exist in the observable Universe, or some fraction of it, must be calculated. To do this, knowledge of or assumptions about both the source and deflector populations must be obtained or made. Secondly, once the idealised strong lens population is characterised, lenses must be simulated and ``observed'' according to the specifications of the target survey. This then determines the number of lenses which will not only exist in a given survey area, but will be detected by the instrument used to observe it. 

Furthermore, it is not straightforward to characterise the number of detectable lenses which will actually end up being identified and catalogued as such. This depends on the strategy used to inspect the data -- for example, expert human visual inspectors versus machine-learning-led searches \citep{Lanusse:2017vha, Marshall2009, More:2024uvq} or citizen science approaches \citep{Marshall:2015fsa, More:2015sfa} -- and the reliability of that strategy; see \cite{DES:2023hxm} for an investigation into the reliability of visual inspection. In turn, the success of an inspection campaign can only be estimated if an accurate forecast for the number of detectable lenses is known. 

This work thus addresses the following question: how well do strong lensing forecasts match what is seen in real data? To answer this question, we make a forecast for the number of detectable lenses in the COSMOS-Web survey and compare their abundance and predicted properties with the first COSMOS-Web catalogue of detected strong lenses, \catname (\catfullname; \citealt{Nightingale2025, Mahler2025}). COSMOS-Web is a survey of $0.54$ deg$^2$ of the COSMOS field, imaged using the NIRCam and MIRI instruments onboard the \textit{James Webb} Space Telescope (\textit{JWST}; \citealt{Casey:2022amu}). The \catname catalogue was constructed following expert visual inspection of NIRCam data, the MIRI observations being omitted at this stage. Accordingly, our forecast is made for the NIRCam observations of COSMOS-Web alone.

The forecasting tool we use is the Python package \lenspop, which was used to forecast strong lens abundances and guide observing strategies for surveys such as \textit{Euclid} and LSST \citep{Collett:2015roa, Mellier2024}. The package was shown to correctly predict the number of strong lenses found in the Strong Lensing Legacy Survey (SL2S; \citealt{Cabanac2007}), along with the distribution of their Einstein radii, validating the forecasting procedure. The package has since been used to forecast the strong lensing abundance for other forthcoming surveys, such as the \textit{Nancy Grace Roman} Space Telescope \citep{Weiner2020} and the \textit{China} Space Station Telescope \citep{Cao:2023bnl}, as well as the abundance of strongly lensed supernovae in LSST observations \citep{Sainz2024}.

Given the demonstrated accuracy of the \lenspop model, we can use the results of the forecast to both interpret current data and guide future strong lensing searches in COSMOS-Web. We further compare our results with the COSMOS-Web strong lensing forecasts made by \cite{Holloway:2023axl} and \cite{Ferrami:2024obm}, which followed different methodologies, as we will later discuss.

This paper is organised as follows: in \autoref{sec:SLinCWeb} we describe the search for strong lenses in COSMOS-Web and the first catalogue of detections. In \autoref{sec:method} we describe our forecasting methodology. In \autoref{sec:results} we present the results of our forecast, whilst in \autoref{sec:discussion} we compare the results of the forecast with the observed data and other forecasts; we conclude with \autoref{sec:conclusions}.

\section{COSMOS-Web \& the search for strong lenses}\label{sec:SLinCWeb}
In this section, we firstly provide an overview of the COSMOS-Web survey, followed by a description of the search for strong lenses in COSMOS-Web.

\subsection{COSMOS-Web}
COSMOS-Web is a survey of a $0.54$ deg$^2$ area of the COSMOS field in the F115W, F150W, F277W and F444W filters of NIRCam, with a $0.19$ deg$^2$ sub-area also imaged with the MIRI instrument in the F770W filter \citep{Casey:2022amu}. The survey design was motivated by previous efforts with the \textit{Hubble} Space Telescope (\textit{HST}) to image certain areas of sky to as great a depth as possible \citep{williams1996, Beckwith2006, Koekemoer2007}, allowing systematic study of galaxies at high redshift; and also by the Cosmic Evolution Survey (COSMOS; \citealt{Scoville:2006vq}) which mapped a contiguous $2$ deg$^2$ sky area, far larger than all previous deep field studies, allowing galaxy formation and evolution to be connected to their environment and the large-scale structure over a vast redshift range, $0 < z < 6$. 

COSMOS-Web emulates that philosophy, with a sky area sufficiently sizeable to capture structures on the order of $100$ Mpc, and imaging reaching $5\sigma$ point-source depths of up to magnitude $28.2$, giving it the ability to detect and observe thousands of galaxies up to $z \sim 11$ \citep{Casey2024, Franco2024} as well as probing both their immediate and wider surroundings. The primary science goals of the survey are the discovery of galaxies in the epoch of reionisation, i.e. $6 \lesssim z \lesssim 11$, as well as mapping the distribution of reionisation and investigating what drives it; the identification of rare quiescent galaxies at $z > 4$ which will allow constraints to be placed on the formation of very massive galaxies; and the measurement of the evolution of the stellar to halo mass relation in galaxies via weak gravitational lensing. 

Furthermore, previous studies of the COSMOS field have yielded the discovery of over 100 strong gravitational lens candidates \citep{Faure:2008dt, Jackson:2008my, More2012, Jin2018, Pourrahmani2018, Garvin:2022gaq, Pearson:2023kfx, Jin2024, Mercier2024}. Thanks to COSMOS-Web, the known strong lens candidates in the field now have high-resolution imaging in the four NIRCam filters in addition to the previous photometric and spectroscopic observations. At the survey's inception, it was estimated that 90 new lenses would be discovered, due to the ability of \textit{JWST} to resolve small Einstein radii lenses and the unprecedented depth of the survey in the near-infrared \citep{Casey:2022amu}. This work refines this estimation, and furthermore, provides predictions for the distributions of lens and source properties, which we can then compare with those of the newly discovered candidate lenses. 

\subsection{The COSMOS-Web Lens Survey}
The initial search for strong lenses in COSMOS-Web was conducted by members of the collaboration's lensing working group, who individually visually inspected images of the $\sim43,000$ most massive galaxies in the COSMOS-Web data as identified as those brighter than AB magnitude 23 in the F277W band, ranking each image as either ``yes'' (the image contains a strong lens), ``maybe'' (the image maybe contains a strong lens), or ``no'' (the image does not contain a strong lens). Visual inspectors were presented with two colour composite images of each candidate, the same images with the lens light subtracted using a S\'ersic light profile, and the S\'ersic profiles used for the subtraction. There were \spectacularlenses candidates that received ``yes'' ratings by more than half the inspectors at this stage, and are considered to be the most spectacular strong lenses in the COSMOS-Web data. A detailed description of these \spectacularlenses objects is presented in \cite{Mahler2025}. 

Over $1,000$ images were ranked as ``maybe'', and of these candidates, all that were ranked as ``yes'' or ``maybe'' by at least half the visual inspection team, plus a significant number of edge cases identified by the lead inspector were selected for detailed follow-up. The \texttt{PyAutoLens}~\href{https://github.com/Jammy2211/PyAutoLens}{\faGithub} software \citep{pyautolens} was used to try to fit a lens model to each of the resulting 419 candidates. The images were then re-examined by six visual inspectors using these lens models, along with a more detailed ranking system. Candidates were ranked as either:
\begin{itemize}
    \item A: High confidence this is a strong lens.
    \item B: Likely a strong lens, but there is ambiguity.
    \item U: Unlikely to be a strong lens, but not impossible.
    \item S: A singly imaged strong lens feature/arc (i.e. without an observed counter image).
    \item X: Not a lens. 
    \item I (Optional and independent): The candidate is interesting and warrants follow-up investigation. 
\end{itemize}
Once each candidate had been graded, it was assigned a score, with 2 points for an A grade and 1 point for either B or S. The highest possible score is therefore 12, for a lens ranked as A by all six inspectors. This procedure led to a discovery of \threequarterlenses strong lenses that scored at least 8/12; there were \halflenses strong lenses that scored at least 6/12. The lens modelling performed by \cite{Nightingale2025} means that all of the candidates have an accompanying lens model; full details of the strong lens search and the resulting catalogue of lenses can be found in that paper.

To facilitate the comparison with the forecast performed in this work, we label the \Ntotal spectacular plus highly ranked candidates (\spectacularlenses + \threequarterlenses) as ``high confidence'' strong lenses, and the \Ntotalweak spectacular plus highly and averagely ranked candidates (\spectacularlenses + \halflenses) as ``high and medium confidence'' strong lenses. For the purposes of this paper, we do not take into account any of the COWLS lens candidates with scores of less than 6/12. We will discuss this, and other selection effects further in \autoref{subsec:selection}.

\section{Forecasting methodology} \label{sec:method}
We use the Python package \lenspop\footnote{\url{https://github.com/tcollett/LensPop}.} to perform our forecast. The package works by computing the idealised number of strong lenses in the Universe, modelling all deflectors as singular isothermal ellipsoids (SIEs) and using galaxies drawn from a given mock catalogue as sources (\lenspop originally contained one such catalogue, but we have modified it to supply our own). The idealised lenses are then ``observed'' according to the chosen survey, yielding not only a prediction for the number of lenses, but also a catalogue of their properties, such as Einstein radii and redshifts. In this section we provide an overview of the \lenspop model, and refer readers to  \cite{Collett:2015roa} for more details. We also describe our modifications to \lenspop which allow us to use it to forecast for COSMOS-Web\footnote{The modified version of \lenspop is available on Github: \url{https://github.com/nataliehogg/lenspop}.}.

\subsection{The deflector population in \lenspop} \label{subsec:lenspopdeflectors}
The \lenspop package simulates a population of deflectors by assuming that the strong lensing cross-section is dominated by elliptical galaxies, meaning that all deflectors are modelled by the SIE profile \citep{Kormann1994}. \nht{The SIE convergence is given by
\begin{equation}
    \kappa(x, y) = \frac12 \left(\frac{\theta_{\rm E}}{\sqrt{qx^2 + y^2/q}}\right),
\end{equation}
where $\theta_{\rm E}$ is the circularised Einstein radius of the lens, $q$ is the axis ratio of the ellipse, and the $x,y$ coordinates are defined in a coordinate system aligned with the major and minor axes of the lens.}
% The SIE density is given by
% \begin{equation}
%     \rho(r) = \frac{\sigma_{\rm v}^2}{2 \pi G r^2},
% \end{equation}
% where $\sigma_{\rm v}$ is the velocity dispersion and $r^2 = qx^2 + y^2/q$ is the elliptical distance from the centre of the profile, $q$ being the axis ratio of the ellipse and $x$ and $y$ being defined in a coordinate system aligned with the major and minor axes of the ellipse. 

The number density of deflectors is computed by integrating a \nht{modified} Schechter function expressed in terms of the velocity dispersion, \nht{$\sigma_{\rm v}$}, derived by \cite{Choi2007} from the Sloan Digital Sky Survey (SDSS) DR5 data \citep{SDSS:2000hjo, SDSS:2007aih},
\begin{equation}
    \mathrm{d}N = \phi_* \left(\frac{\sigma_{\rm v}}{\sigma_*}\right)^\alpha \exp \left[- \left(\frac{\sigma_{\rm v}}{\sigma_*}\right)^\beta\right] \frac{\beta}{\Gamma(\alpha/\beta)} \frac{1}{\sigma_{\rm v}} \mathrm{d} \sigma_{\rm v}, \label{eq:veldisp}
\end{equation}
where $\phi_* = 8 \times 10^{-3} h^3$Mpc$^{-3}$, $\sigma_* = 161$ kms$^{-1}$, $\alpha=2.32$ and $\beta=2.67$. \nht{Here, $\alpha$ describes the low-velocity power law index, and $\beta$ describes the high velocity exponential cut-off index.} The shape and normalisation of this function do not evolve with redshift.
% see also eqn 1 of https://arxiv.org/pdf/astro-ph/9806080

Velocity dispersions for each deflector are drawn from \autoref{eq:veldisp}, whilst redshifts for each deflector are drawn from the differential comoving volume defined by the maximum deflector redshift. In the original \lenspop model this was set at $z=2$, but in this work we increase the maximum redshift to $z=3$, since we expect that higher redshift lenses will be present in COSMOS-Web due to \textit{JWST}'s sensitivity to faint and red objects. For example, the first strong lens detected in the survey, the COSMOS-Web~Ring, is at $z=2.02$, with a source at $z=5.10$ \citep{Mercier2024, Shuntov2025}. 

Since all deflectors are modelled as SIEs\footnote{Above $z=3$, we may expect that a larger proportion of the deflector population will be disk, spiral or irregular galaxies, necessitating a modification of the deflector model to account for this. We leave the quantification of a potential $z\geq3$ deflector population to a future work.}, their Einstein radii are given by
\begin{equation}
    \theta_{\rm E} = 4 \pi \frac{\sigma_{\rm v}^2}{c^2} \frac{D_{\rm ds}}{D_{\rm os}}, \label{eq:thetaE}
\end{equation}
where $D_{\rm ds}$ and $D_{\rm os}$ are the angular diameter distances between deflector and source and observer and source respectively. These are computed from the redshift for each deflector and source (once the source population has been drawn, see \autoref{subsec:lenspopsources}), in a spatially flat $\Lambda$CDM cosmology with $\Omega_{\rm m} = 0.3$, $\Omega_{\Lambda} = 0.7$ and 
% $H_0 = 70$~kms$^{-1}$ Mpc$^{-1}$. 
$h = 0.7$.
The axis ratio for each deflector is drawn from a Rayleigh distribution, truncated at $q=0.2$ to avoid highly elliptical mass profiles.

The deflector light profiles are modelled by an elliptical de Vaucouleurs profile (\cite{deVaucouleurs1948}; equivalent to the S\'ersic profile with the S\'ersic index $n=4$) whose centre and ellipticity are the same as those of the mass profile. Given a velocity dispersion, the fundamental plane relation for elliptical galaxies yields the effective radius and absolute magnitude for the deflector's light. The \lenspop model employs the fundamental plane obtained  by \cite {Hyde:2008yf, Hyde:2008yh} from SDSS DR4 and DR6 data.
% \begin{equation}
%     \log_{10}\left(\frac{R_{\rm eff}}{\rm kpc}\right) = a \log_{10} \left(\frac{\sigma_{\rm v}}{\mathrm{kms}^{-1}}\right) + b \frac{\mu_{\rm eff}}{\mathrm{mag}} + c,
% \end{equation}
% In the \lenspop model, the effective radius of the light profile is the same in all bands whilst the absolute magnitude can vary across bands. 

The quantities that describe each deflector are therefore its redshift $z_{\rm lens}$, velocity dispersion $\sigma_{\rm v}$, Einstein radius $\theta_{\rm E}$, axis ratio $q$, effective radius $R_{\rm eff}$ and absolute magnitude $M$ in each band. Weak lensing by line-of-sight objects is not included in the model. We do not modify any part of the above computation for our work.

\subsection{The source population in \lenspop} \label{subsec:lenspopsources}

The source population parameters are the redshifts, light profiles, and the absolute magnitudes in each observation band. The redshifts and magnitudes are drawn from a given mock galaxy catalogue. In the original \lenspop model, the mock galaxy catalogue used was produced for LSST by adding baryonic effects such as gas cooling, star formation, supernovae and active galactic nuclei to the dark-matter-only Millennium simulation \citep{DeLucia:2005yhi, Connolly2010}. The catalogue covers a $4.5 \times 4.5$ deg$^2$ area of the sky, with the highest redshift objects being around $z=6$ and the smallest halo mass being $2.5 \times 10^9$ M$_{\odot}$.

The source light profiles are all assumed to be elliptical exponential profiles, with the ellipticity drawn from a Rayleigh distribution truncated at $q=0.2$, in line with the ellipticity distribution of the deflector profiles. The effective radii of the source light profiles are given by
\begin{equation}
    \log_{10}\left(\frac{R_{\rm eff}}{\rm kpc}\right) =  \left(\frac{M}{-19.5}\right)^{-0.22} \left(\frac{1 + z}{5}\right)^{-1.2} + \mathcal{N},
\end{equation}
where $M$ is the absolute magnitude and $\mathcal{N}$ is a normally distributed random variable with a scatter of $0.3$.

To improve the suitability of \lenspop as a forecasting tool for COSMOS-Web, we must modify the source population used. Since \textit{JWST} is a near-infrared telescope, it is sensitive to fainter and redder sources than those in the surveys that \lenspop was originally designed to forecast for, such as LSST. To ensure that these faint sources are present in our forecast, we added a new mock galaxy catalogue to \lenspop to be used as the source population. 

We chose to use the JAdes extraGalactic Ultradeep Artificial Realization (JAGUAR) mock catalogue \citep{Williams2018}, which was originally created during the survey design of the \textit{JWST} Advanced Deep Extragalactic Survey (JADES), a Guaranteed Time Observation program to observe the Hubble Deep and Ultra Deep fields using all three \textit{JWST} instruments: NIRCam, NIRSpec and MIRI \citep{Eisenstein2023}.

The final JAGUAR mock catalogue comprises ten realisations of an $11 \times 11$ arcmin$^2$ area, with each realisation containing roughly 300,000 objects up to $z \sim 15$, with the smallest halo mass being $1\times10^{6}$ M$_{\odot}$ \citep{Williams2018}. The catalogue contains both star-forming and quiescent galaxies. A spectral energy distribution (SED) is assigned to each mock galaxy using the \texttt{BEAGLE} tool \citep{Chevallard2016}, with the complete catalogue providing galaxy properties such as redshift, half-light radius, axis ratio and fluxes corresponding to observations in NIRCam filters. To ensure consistency in the properties of our source population, we replace the calculations of the effective radius and axis ratio for each source described above with their values provided in the catalogue. We use only a single realisation of the mock. We note that the JAGUAR mock catalogue covers a relatively small area of the sky; the main effect this will have on our results is a potential underestimation of the abundance of the most massive source galaxies in the Universe.

\subsection{Observing and detecting lenses} \label{subsec:lenses}

After the creation of the deflector and source populations following the procedure described above, the number of strong lenses potentially observable in the sky can be computed. Firstly, \autoref{eq:veldisp} is integrated with respect to redshift, which provides the number of galaxies which can act as deflectors. % gave up on trying to write nice optical depth integrals; see drawSourcePopulation in PopulationFunctions.py
Whether a given deflector lenses a source is computed in the following manner: a square box around each deflector is constructed such that it contains, on average, one source galaxy. This is done using the source area number density, $d$, such that the length of the side of the box in arcseconds is given by $l=\sqrt{1/d}$. For the source catalogue used in this work, $l \approx 1.2''$. The deflector is located at the centre of the box, whilst the source is placed within the box at random. The source--deflector pair thus create a strong lensing system if and only if $z_{\rm source} > z_{\rm lens}$ and $\theta_{\rm E}^2 > x_{\rm s}^2 + y_{\rm s}^2$, i.e. the source is located within the Einstein radius of the deflector in the image plane. We note that if (large) external shear induced by line-of-sight objects were to be included in the lens model the latter criterion would no longer necessarily hold, as the astroid caustics characteristic of shear can extend outside the radial caustic associated with the SIE potential, altering the strong lensing cross-section \citep{Kochanek:2004ua, Huterer:2004jh, Lee:2014yxa}.

In the original \lenspop model, this calculation led to an ideal population of $1.1\times 10^{7}$ lenses; with our modifications, we find $2.8\times 10^8$. This order of magnitude increase comes from the greater depth of the new source population compared to the old, as well as its increased redshift range. We also simulate deflectors up to $z=3$, rather than $z=2$ in the original work of \cite{Collett:2015roa}. 

At this point, the  so-called transfer function is applied to determine which strong lenses from the idealised population could be observed by a given survey. To do this, the idealised lenses are evaluated on a pixel grid, with a pixel scale given by the pixel scale of the chosen survey, and convolved with a Gaussian point spread function (PSF). Poisson noise is added, and further specifications of the survey are taken into account, including sky area, seeing (for ground-based telescopes), exposure time, zero-points and filters. For the complete survey specifications used in the original \lenspop model, see Table 1 of \cite{Collett:2015roa}.

For this work,  we added the filter transmission data \citep{Rodrigo2012} for the four NIRCam filters used in COSMOS-Web to \lenspop, along with the survey specifications for COSMOS-Web; see \autoref{tab:specs} for the specifications used. Note that we apply the same pixel scale of $0.031$ arcseconds to all four bands, whereas in reality the F277W and F444W bands have a pixel scale of $0.063$ arcseconds. This is because \texttt{lenspop} currently only allows for a single value to be passed as the pixel scale. 

\begin{table}
    \centering
    \begin{tabular}{Slcccc}
    Property & \multicolumn{4}{c}{Value(s)} \\
    \hline
    \hline
    Sky area [deg$^2$]    & \multicolumn{4}{c}{0.54} \\
    Exposure time [s] & \multicolumn{4}{c}{257} \\
    Pixel scale [$^{\prime\prime}$] & \multicolumn{4}{c}{0.031}\\
    Filters & F115W & F150W & F277W & F444W \\
    PSF FWHM [$^{\prime\prime}$] & 0.04 & 0.05 & 0.092 & 0.145 \\ %0.07 arcseconds \\
    Sky brightness [mag/arcsec$^{2}$] & 30.96 & 29.96 & 28.96 & 28.15 \\
    \hline\\
    \end{tabular}
    \caption{Properties of the COSMOS-Web survey used in our \lenspop forecast. }
    \label{tab:specs}
\end{table}

Furthermore, we do not modify the PSF computation, with each image still being convolved with a Gaussian PSF. We do not expect this to significantly impact the forecast number of detectable lenses, as the dominant PSF-related selection effect for extended sources is that of its size (full width at half maximum; FWHM) rather than its shape \citep{Sonnenfeld2023b}. We change the default PSF FWHM used in \lenspop to those of the NIRCam filters; the values are listed in \autoref{tab:specs}.

Lastly, for a lens to be deemed detectable, it must pass a number of selection cuts. The criteria specified by \cite{Collett:2015roa}, which we also follow, are:
\begin{enumerate}
    \item Multiple images of the source must be produced, $\theta_{\rm E}^2 > x_{\rm s}^2 + y_{\rm s}^2$;
    \item The image and counter-image must be resolved, $\theta_{\rm E}^2 > r_{\rm s}^2 + (s/2)^2$, where $r_{\rm s}$ is the angular unlensed source size and $s$ is the seeing (PSF FWHM);
    \item Tangential shearing of the arcs must be detectable, $\mu_{\rm tot} r_{\rm s}> s$, $\mu_{\rm tot} > 3$, where $\mu_{\rm tot}$ is the total magnification of the source;
    \item The signal-to-noise ratio must be large enough that the previous criteria are obviously satisfied to a human visual inspector, $\mathrm{SNR}_{\rm tot}> 20$, where SNR$_{\rm tot}$ is the total signal-to-noise ratio.
\end{enumerate}

\section{Results} \label{sec:results}
In this section, we describe our results. We note that the forecast presented in this work was produced by NBH concurrently with the first round of visual inspection of the data, and there was no attempt to keep the forecasted results blinded from the inspection team whilst the visual inspection was ongoing. NBH also contributed to both rounds of visual inspection described in \cite{Mahler2025} and \cite{Nightingale2025}.

\subsection{The forecast abundance of strong lenses}
With the above described modifications to \lenspop, we predict that 107 strong lenses (i.e. objects that meet the four criteria outlined in \autoref{subsec:lenses}) will be detected in COSMOS-Web across all bands, assuming lens light subtraction. To estimate a lower bound on the number of detectable lenses, we can examine so-called difference images, produced by subtracting the data in a given band from another at a shorter wavelength \citep{Gavazzi:2014nza}. We compute the F115W$-$F150W, F115W$-$F277W and F115W$-$F444W differences, finding that 14 lenses would be discovered in each.

Our result compares favourably with the number of strong lenses found in COSMOS-Web after two rounds of visual inspection. After the first round of visual inspection, \spectacularlenses strong lenses were discovered \citep{Mahler2025}, which already exceeds the lower bound on the abundance predicted in our difference images. The second round of visual inspection targeted the 419 most promising remaining candidates; based on their scores, we identify \Ntotal as high confidence strong lenses, or \Ntotalweak as high and medium confidence strong lenses.

Considering just the number of high confidence strong lenses, and assuming the \lenspop model is  perfectly accurate, this implies that potentially as many as 48 strong lenses are present in the COSMOS-Web field that were not identified as such by the \cite{Mahler2025} and \cite{Nightingale2025} visual inspections, or that some of the weaker candidates from those inspections should have their scores reassessed and increased. These may only be identified after spectroscopic follow-up to obtain lens and source redshifts. On the other hand, if instead the visual inspection is considered to be perfectly accurate and that all true strong lenses in COSMOS-Web are already included in the high confidence sample, this implies that the \lenspop model is significantly over-predicting the abundance of strong lenses.

With the small number statistics of COSMOS-Web, this difference between prediction and reality may be considered tolerable, but for larger surveys such as \textit{Euclid}, a detection of up to 50\% fewer high confidence candidate lenses than predicted by \cite{Collett:2015roa} may be dissatisfying. It is also worth noting that the first systematic search for strong lenses in \textit{Euclid} Early Release Observations, which covered $0.7$~deg$^2$ of the sky, led to the discovery of three `A' grade and 13 `B' grade candidates\footnote{The \textit{Euclid} strong lens gradings are slightly more confident than those defined in this paper, with `A' grade candidates considered to be ``sure lenses'' and `B' grade candidates considered ``probable lenses''.}, of which five were successfully fit with a lens model \citep{Euclid:2024jyk}. Extrapolating to the full \textit{Euclid} area, this implies $100,000^{+70,000}_{-30,000}$ strong lenses. The $170,000$ strong lenses predicted by \cite{Collett:2015roa} with \lenspop sits on the upper limit of this extrapolation.

\subsection{Simulated images} \label{subsec:simages}

\begin{figure*}
    \centering
    \includegraphics[width=0.98\textwidth]{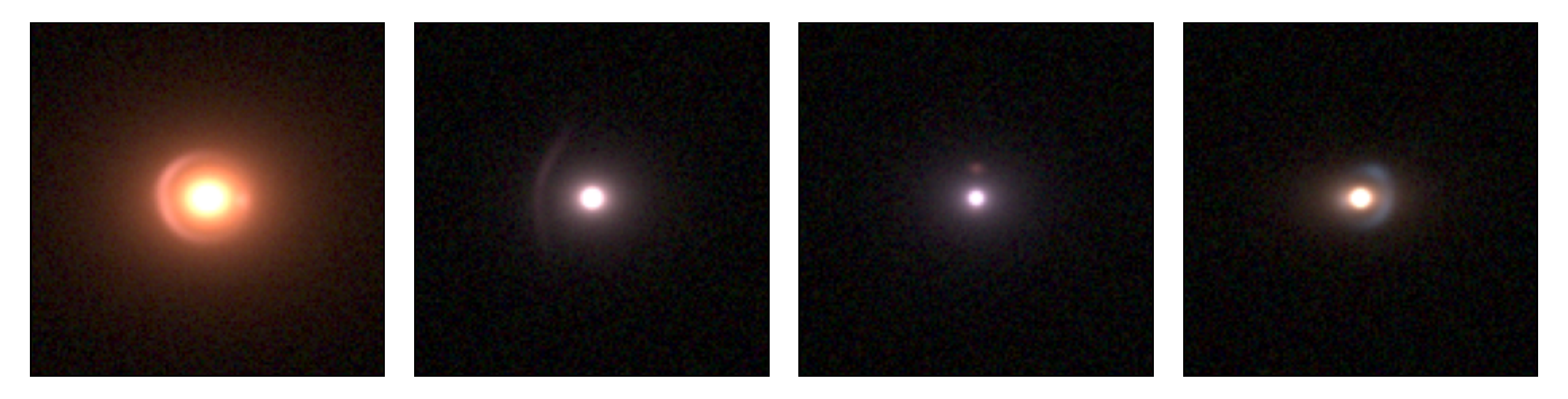}
    \caption{Multiband images of four simulated COSMOS-Web-like lenses using properties of the forecast population. From left to right, there is a near-complete Einstein ring, an extended arc, a single image (the counter-image being obscured by the lens light) and a distinctive blue lensed source. The colours are representative and are produced by combining the data simulated in three different NIRCam bands (F150W, F277W and F444W).}
    \label{fig:long_grid}
\end{figure*}

The lens properties in our forecast catalogue can be used to simulate images of COSMOS-Web-like strong lenses. Four illustrative examples are shown in \autoref{fig:long_grid}, produced using the \texttt{lenstronomy}~\href{https://github.com/lenstronomy/lenstronomy}{\faGithub} package \citep{Birrer:2018xgm, Birrer2021}. Each panel shows a type of strong lens that could be seen in COSMOS-Web, with different lens and source colours clearly visible. The colours are representative and are produced by combining the data simulated in three different NIRCam bands, F150W, F277W and F444W. The images follow the same scale in magnitude and are ordered from left to right in terms of decreasing magnitude of the lens light in the F444W band.

As is typical with strong lensing, the lensed source light in each case is faint compared to the light of the lens galaxy, and the small Einstein radii mean that some lensed features are partially obscured by or blended with the lens light. Since the \lenspop model assumes all deflectors are SIEs, the arc features in the simulated images are regular and easily identifiable as the product of strong lensing. This is not the case in real data, where deflectors can have more complex gravitational potentials, which, combined with complex lens and source light morphology, can lead to less obvious strong lensing features. All of these problems were motivators for the second stage of visual inspection by \cite{Nightingale2025}, in which lens models were fit to the candidate lenses and inspected along with the images as an additional discriminatory factor. 

\subsection{Properties of the forecast strong lenses} \label{subsec:properties}

\begin{figure*}
    \centering
    \includegraphics[width=0.98\textwidth]{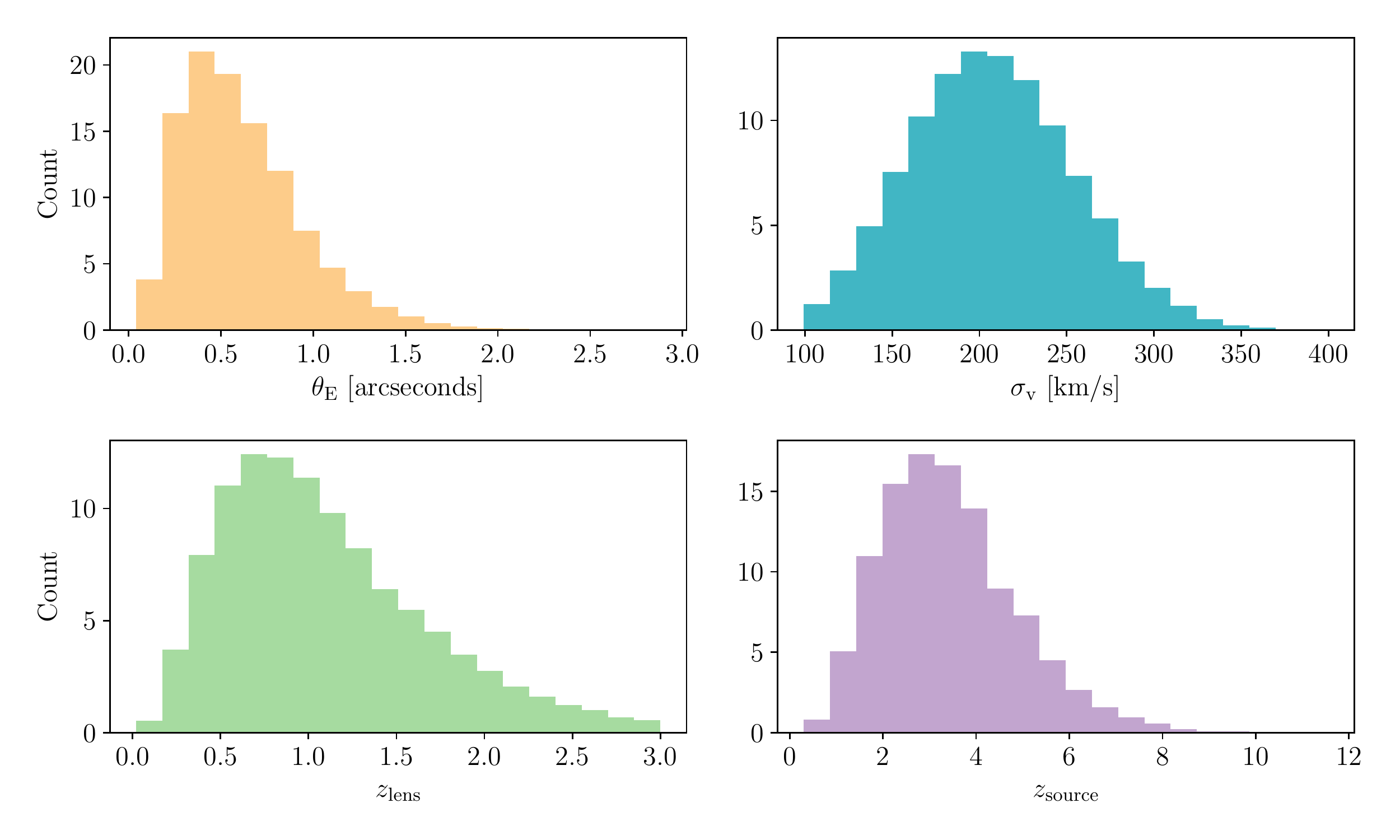}
    \caption{The distribution of Einstein radii (top left), velocity dispersions (top right), lens redshifts (bottom left) and source redshifts (bottom right) for our forecast lenses. Each histogram is weighted so the total count sums to 107, the number of lenses we forecast for COSMOS-Web.}
    \label{fig:thetaEzhist}
\end{figure*}

Along with the calculation of the number of strong lenses in the survey, we can also examine the properties of the forecast systems. In \autoref{fig:thetaEzhist}, we show the distributions of Einstein radii, velocity dispersions and lens and source redshifts for our forecast catalogues. The median Einstein radius is $0.55$ arcseconds, with $88\%$ of the lenses having Einstein radii smaller than $1$ arcsecond. The PSF in the F115W band provides a hard lower limit on the Einstein radius of a detectable lens, 0.04 arcseconds. Only $0.2\%$ of the lenses have Einstein radii larger than 2 arcseconds. In conjunction with the distribution of velocity dispersions, which peaks at $200$ kms$^{-1}$, this implies that the majority of strong lenses detected in COSMOS-Web (and by extrapolation, in other small area \textit{JWST} surveys) will be low-mass galaxies.

 % In \autoref{fig:diff_hist}, we overplot the distribution for the velocity dispersion of lenses detectable in the F115W$-$F444W difference images on the original distribution for lenses detectable in a single band. This shows that the lenses detectable in the difference images tend to have more massive deflectors than those detectable in a single band.  This is consistent with the findings of \cite{Collett:2015roa} for the Dark Energy Survey (DES) and LSST lenses.

% \begin{figure}
%     \centering
%     \includegraphics[width=0.47\textwidth]{plots/diff_hist_sigv.pdf}
%     \caption{Distributions of the velocity dispersions in our COSMOS-Web-like forecast systems. The coloured histogram represents lenses detectable in any single band, whilst the black histogram represents lenses detectable in the F115W$-$F444W difference images.}
%     \label{fig:diff_hist}
% \end{figure}

\begin{figure}
    \centering
    \includegraphics[width=0.47\textwidth]{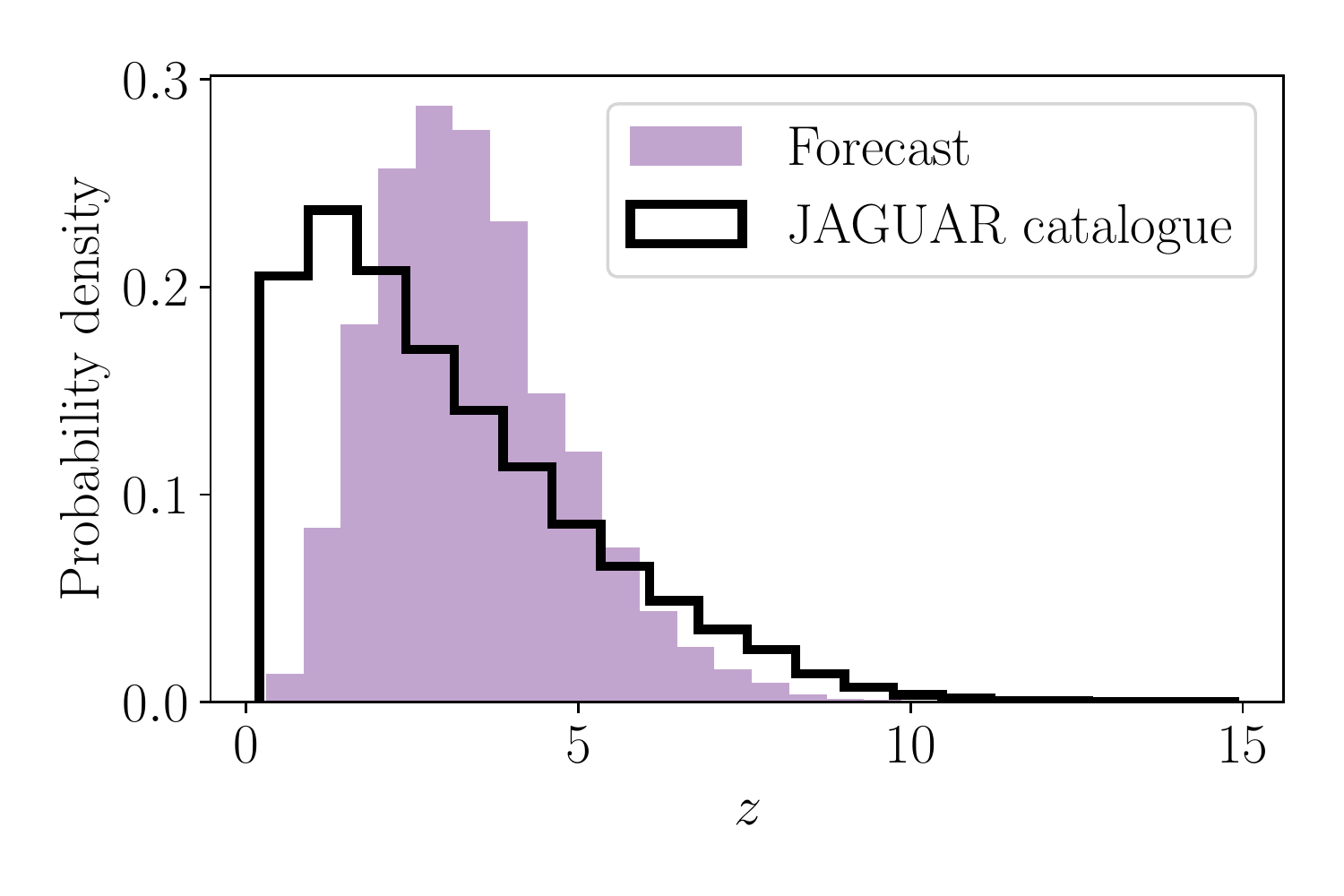}
    \caption{Distribution of the source redshifts in our COSMOS-Web-like forecast systems (purple) and the redshifts of galaxies in the JAGUAR mock catalogue from which the source population was drawn (black).  }
    \label{fig:zdists}
\end{figure}

The bottom left panel of \autoref{fig:thetaEzhist} shows that the lens redshift distribution peaks at $z=1$, with a long tail extending to our maximum allowed lens redshift, $z=3$. The source redshift distribution, shown in the bottom right panel, peaks at $z=3.2$. Despite the source population containing galaxies up to $z\sim 15$, the maximum source redshift in our catalogue of forecast lenses is $z=11.6$, with $5.3\%$ of the sources lying between this value and $z=6$. This explicitly demonstrates the potential that COSMOS-Web and other \textit{JWST} surveys have to find strong lens source galaxies that lie in the epoch of reionisation. The difference between the distributions of the forecast source redshifts and that of the JAGUAR catalogue from which the source population is drawn is shown in \autoref{fig:zdists}. 

The absence of the high-redshift tail in the forecast source redshift distribution implies that lensing of extremely high redshift sources is rare (as expected due to the comparative rarity of galaxies at high redshifts, meaning there are fewer sources to be lensed), and that even if an extremely high redshift source is lensed, the magnification is likely insufficient to lead to a detection in COSMOS-Web. However, a discovery of systems with source redshifts $ 6 \lesssim z \lesssim 11$ in COSMOS-Web data -- possible if our forecast is correct -- would represent the highest redshift lensed sources ever discovered. 

\begin{figure}
	\centering
	\includegraphics[width=0.47\textwidth]{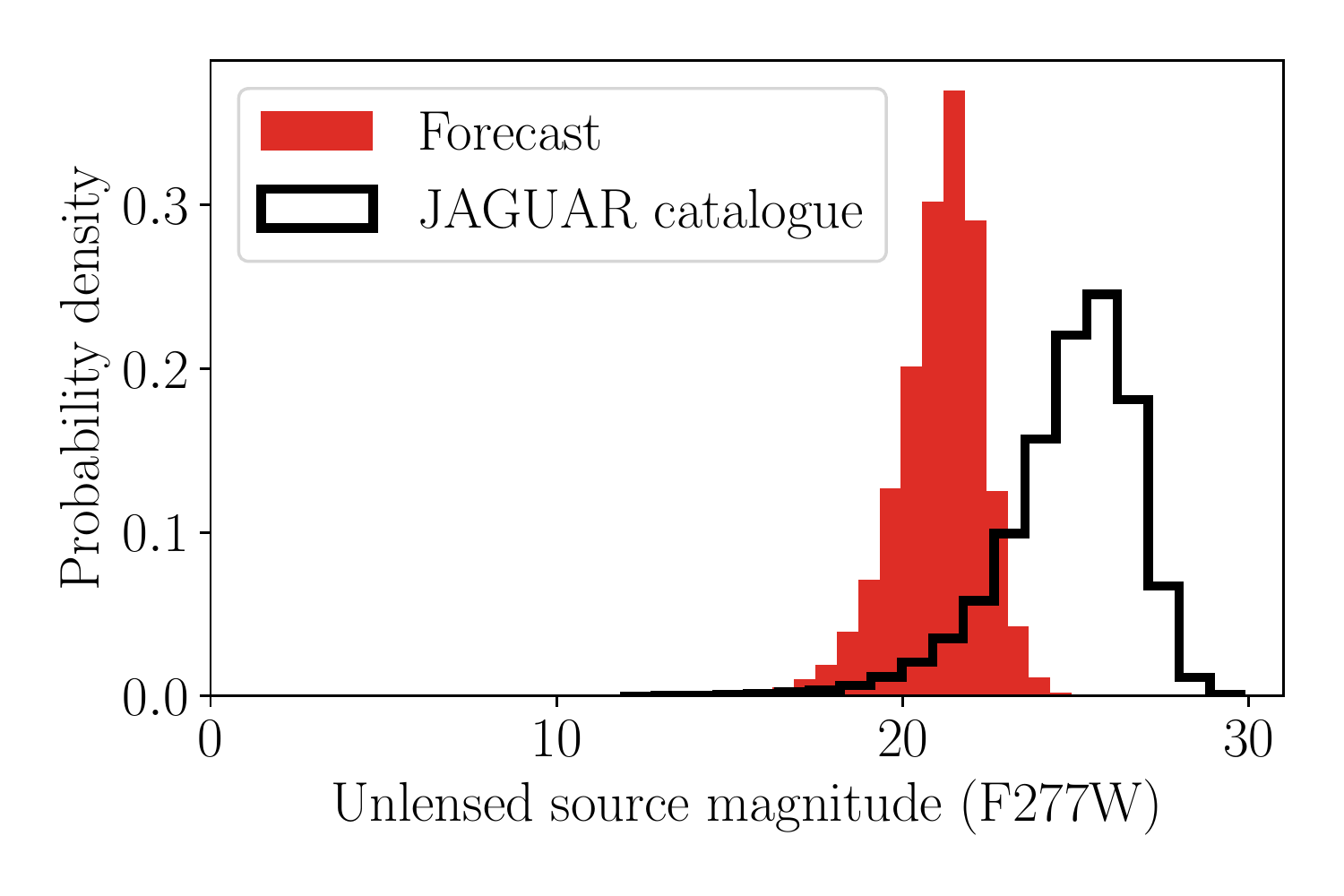}
	\caption{Distribution of the forecast unlensed source magnitude in the F277W band (red) and  the unlensed source magnitude in the JAGUAR catalogue from which the forecast population was drawn (black). }
	\label{fig:sourcemags}
\end{figure}

In \autoref{fig:sourcemags} we show the distribution of the forecast unlensed source AB magnitude in the F277W band, along with the distribution of the equivalent magnitude in the JAGUAR catalogue from which the sources were drawn. We choose the F277W band as a representative example of the distribution of source magnitude in each of the four bands. From this figure we can see that the majority of the sources in the forecast are drawn from the brightest sources in the JAGUAR catalogue, meaning that the forecast population is not underestimated due to lack of depth in the JAGUAR catalogue. We can also compare our forecast to the distribution of source magnitudes in the LSST catalogue used for the source population by \cite{Collett:2015roa}. For example, in the $z$-band of the LSST catalogue, $2.0\%$ of sources have magnitudes fainter than 28, whereas in our forecast for the F115W filter of NIRCam\footnote{We select the LSST $z$-band  for direct comparison to our forecast as it is the closest in wavelength to the F115W filter of NIRCam.}, $6.8\%$ of sources are fainter than magnitude 28. This highlights the importance of our use of the JAGUAR catalogue instead of the LSST catalogue of the original \lenspop forecasts, as without it we would have possibly missed the faintest sources discoverable in COSMOS-Web.

The comparative lack of lenses with large Einstein radii in our forecast is correlated with the features of the other distributions described above. The lenses expected to be detected in COSMOS-Web are low-mass galaxies, with $8.3\%$ occurring above $z=2$. The Einstein radius of an isothermal lens is proportional to $D_{\rm ds}/ D_{\rm os}$ (see \autoref{eq:thetaE}), meaning that for a given source redshift, high-redshift lenses have smaller Einstein radii. Given that large Einstein radii lenses are typically easy to identify via visual inspection, they are unlikely to have been missed in the COSMOS-Web data despite their expected rarity. As we will discuss further in \autoref{subsec:datacomparison}, this expectation is supported by the distribution of the observed Einstein radii of the \spectacularlenses spectacular lenses, which is skewed towards larger values.

\section{Discussion} \label{sec:discussion}

\subsection{Comparison with properties of the observed data} \label{subsec:datacomparison}
Besides the comparison of the abundance of strong lensing in COSMOS-Web with our forecast, the lens modelling performed by \cite{Nightingale2025} allows us to compare properties of our forecast lenses with properties of the observed strong lenses. 
\subsubsection{Einstein radii}

\begin{figure*}
	\centering
	\includegraphics[width=\textwidth]{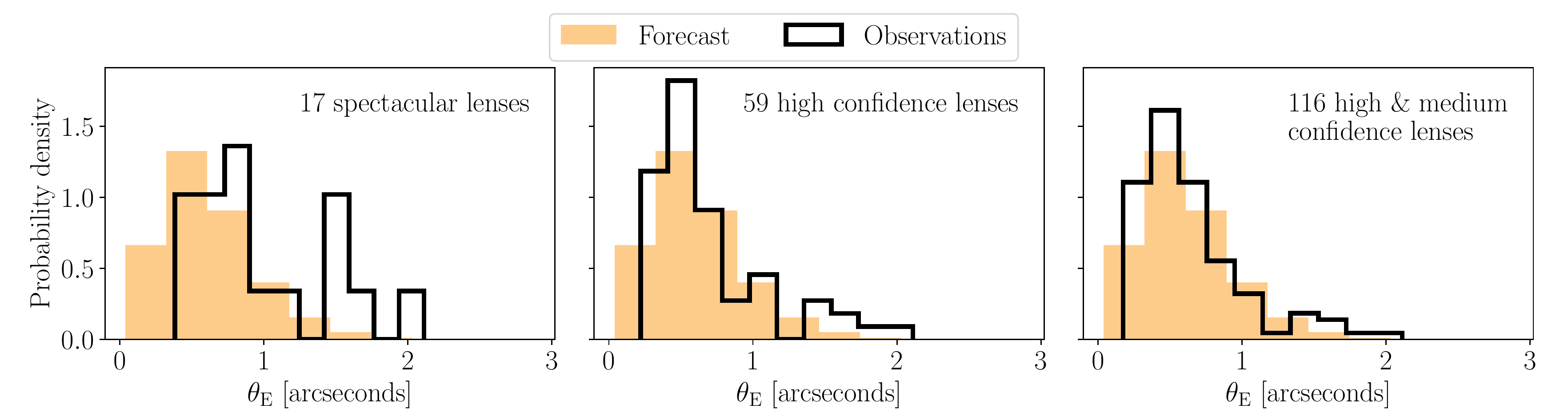}
	\caption{Our forecast distribution of Einstein radii (yellow) compared to the observed lenses (black). \textit{Left}: the \spectacularlenses spectacular lenses. \textit{Middle}: the \Ntotal high confidence lenses. \textit{Right}: the \Ntotalweak high and medium confidence lenses.  }
	\label{fig:thetaEcomparison}
\end{figure*}

In \autoref{fig:thetaEcomparison} we show the forecast distribution of Einstein radii (yellow histograms) compared to the distribution of this parameter obtained from modelling the observed strong lenses (black unfilled histograms). Our forecast yields a single value for this parameter, computed using \autoref{eq:thetaE}; the lens modelling of \cite{Nightingale2025} was performed on a single band per candidate lens, using whichever band was judged to have the brightest source light. The Einstein radius was computed as the area within the critical curve for the inferred SIE mass model for the given band. 

\begin{figure*}
	\centering
	\includegraphics[width=\textwidth]{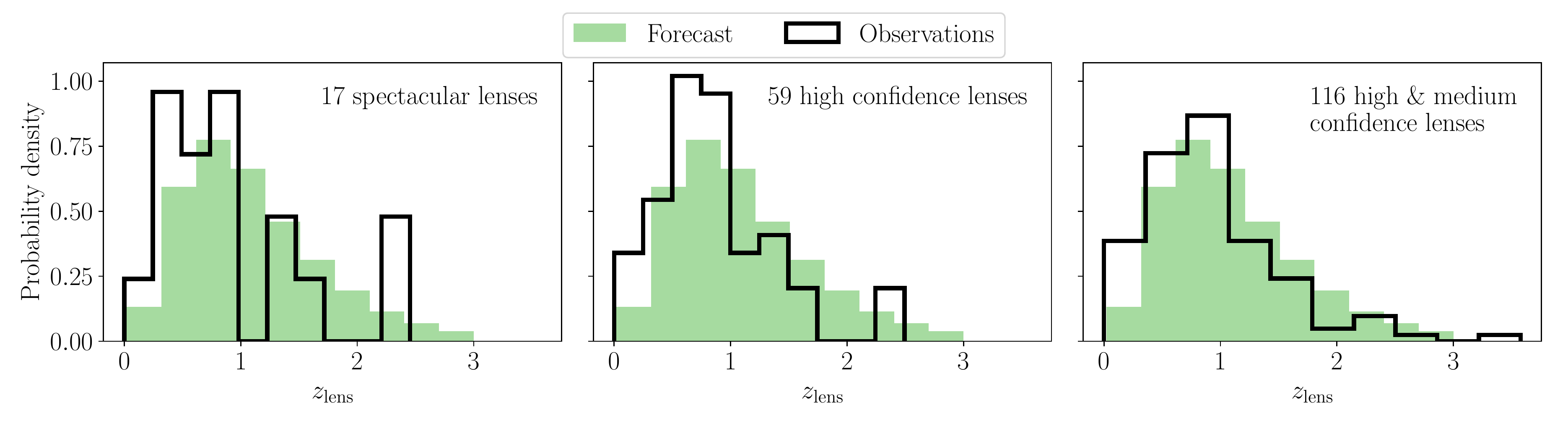}
	\caption{Our forecast distribution of lens redshifts (green) compared to the observed lenses (black). \textit{Left}: the \spectacularlenses spectacular lenses. \textit{Middle}: the \Ntotal high confidence lenses. \textit{Right}: the \Ntotalweak high and medium confidence lenses. For the observed lenses we show the photometric redshifts.  }
	\label{fig:lenszcomparison}
\end{figure*}

From \autoref{fig:thetaEcomparison}, we can see that, as expected, the \spectacularlenses spectacular lenses do not contain any very small Einstein radius systems, and larger values of this quantity are over-represented compared to the forecast distribution. In contrast, once the additional lenses identified in the second round of visual inspection are added, the distributions of the Einstein radii become more similar to the forecast distribution, although the number of very small Einstein radius systems is still less than predicted by our forecast.

Assuming the \lenspop forecast to be perfectly accurate, this result motivates re-inspection of the COSMOS-Web data, with a focus on lower mass galaxies, such as those observed by \cite{Gelli2023, Asada2023, Lin2023},  which will in turn have smaller Einstein radii. \nht{A less stringent magnitude cut in the preparation of the data to be inspected may result in more low-mass galaxies and therefore more lenses with small Einstein radii being identified; though, since many of these Einstein radii are likely to be close in angular size to the PSF FWHM of NIRCam, with the very smallest being unresolved in the longer wavelength bands and approaching the PSF FWHM in the shorter wavelength bands, they will be challenging for visual inspectors to correctly identify \citep{DES:2023hxm}}. However, they may be an ideal target for machine-learning-led searches. The over-representation of larger Einstein radius systems in the observed data compared to the forecast can be understood from the fact that the observed data contains a number of group-scale systems (strong lensing by a group of galaxies), which is not taken into account in the \lenspop model.

We perform the two-sample Kolmogorov--Smirnov (KS) test \citep{Sprent1998} to determine the similarity of the forecast and observed distributions. The two-sample KS test measures the maximum vertical distance between cumulative distribution functions associated with two samples of data. Our null hypothesis is that the observed Einstein radii are drawn from the same distribution as that of our forecast sample. We run the KS test on all three observed datasets: the \spectacularlenses spectacular lenses, the \Ntotal high confidence lenses and the \Ntotalweak high and medium confidence lenses (both of the latter including the \spectacularlenses spectacular lenses). 

We find that the null hypothesis is rejected with 95\% confidence in the case of the spectacular lenses, with a p-value of $0.006$, but is not rejected for the larger samples, with p-values of $0.56$ and $0.78$ respectively. The KS statistic decreases from $0.10$ for the high confidence lenses to $0.06$ for the high plus medium confidence lenses, implying that the distribution of Einstein radii in the \Ntotalweak high plus medium confidence lenses is most similar to that of the forecast distribution.

% watch out bc these tests return NaNs if you have NaNs/blank cells in your csv
A drawback of the KS test is that it is most sensitive to differences in the centre of distributions rather than in the tails \citep{Razali2011}. Since both our forecast and observed distributions of Einstein radius are clearly positively skewed, we can instead use the Anderson--Darling (AD) test \citep{AndersonDarling1952}, which is more sensitive to differences in the tails, to compare the distributions. For the distribution of the Einstein radii of the spectacular lenses, the null hypothesis that this sample is drawn from the same underlying distribution as the forecast sample is rejected at $95\%$ confidence by the AD test, with a p-value of $0.001$. However, as with the KS test, the null hypothesis is \textit{not} rejected at this confidence level for the distributions of the Einstein radius in the high confidence and high and medium confidence lenses, with p-values of $0.16$ and $0.25$ respectively. The AD test statistics for the latter two cases are $0.75$ and $0.036$, implying that, under the AD test, the \Ntotalweak high confidence lenses are also the most similar to the forecast distribution. We summarise all the results of our statistical tests in tables in Appendix~\ref{sec:stattables}.

It must be borne in mind that the overall similarity between the forecast and observed Einstein radius distributions is driven in part by how the Einstein radii are computed; in the forecast, all lenses are assumed to have SIE mass profiles, and all the candidate lenses in the observed data were fit with SIE profiles, meaning that we are comparing like with like. However, future modelling of the candidates may find that more complex mass profiles are preferred for some of the lenses, making comparison with the \lenspop forecast less robust. 

% in the end I'm not sure if this really matters; Einstein radius is Einstein radius, regardless of the mass profile it's come from, no? but then again lenspop forecasts x many SIE lenses, which if you then select based on idk NFW, you are not necessarily going to select the same lenses and ergo you may get a different theta_E dist. so it's probably a small effect but an effect nevertheless

% Furthermore, the lack of `external' shear due to line-of-sight effects in the \lenspop model is another important simplification, given that we know ext shear is very large for most lenses.

\subsubsection{Lens redshifts}

In \autoref{fig:lenszcomparison} we show the distributions of the lens redshifts in our forecast (green histograms) and observed (unfilled black histograms) populations. We use the estimated photometric redshifts of the strong lenses for our comparison; these redshifts have been checked against spectroscopic measurements where available, and are in good agreement with those, albeit with some scatter above $z=1$. We choose to use the photometric redshifts as they extend to $z\sim4$, whilst the spectroscopic redshifts only reach $z\sim2$, below the redshift limit imposed in our forecast. More details on the redshifts of the observed strong lenses can be found in \cite{Nightingale2025}; importantly, there is no correlation between lens candidate ranking and lens redshift. 

From \autoref{fig:lenszcomparison}, we can see that, in general, the predicted distribution does not fully capture any of the observed populations at low redshifts. The observed lens redshift distribution in all cases peaks below $z=1$, and slightly below the peak of the forecast distribution in each case. For the \Ntotalweak high and medium confidence lenses, it is clear that the high-redshift tail of the distribution is more extended in the observed data compared to the forecast. In particular, there is a single lens, COSJ100151+022347, whose photometric redshift is estimated at $z=3.58$, well above the limit we imposed in our forecast of $z=3$. This lens received a grade of 6/12, based on two `A' grades, two `B' grades, one `U' grade and one `X' grade, and does not have a spectroscopic redshift. If the photometric redshift of this and the other five lenses above $z=2$ were to be confirmed with spectroscopy, this would be indicative of the need to re-run the forecast with an even higher upper limit for the lens redshift distribution.

\begin{figure*}
% \labellist
% \small\hair 2pt
%     \pinlabel \textcolor{green}{hello world} at 50 50
% \endlabellist
    \centering
    \includegraphics[width=0.19\textwidth]{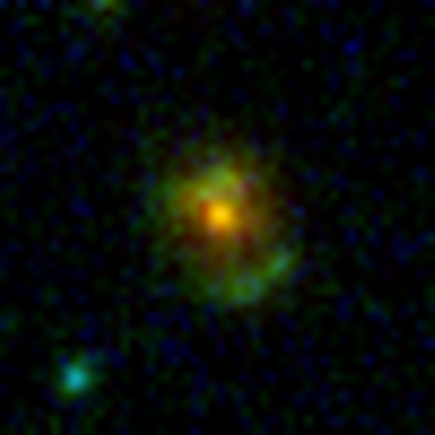}
    \includegraphics[width=0.19\textwidth]{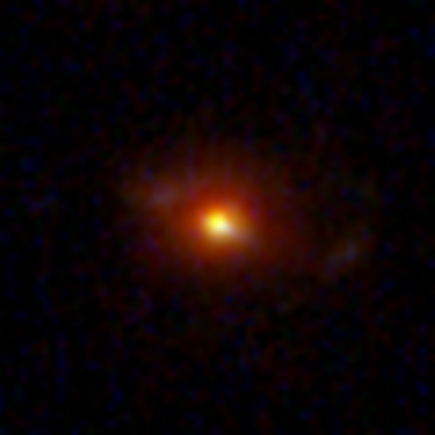}
    \includegraphics[width=0.19\textwidth]{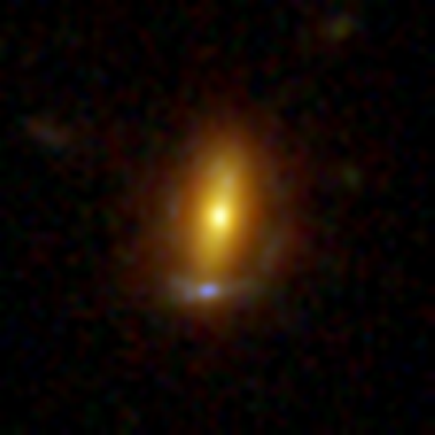}
    \includegraphics[width=0.19\textwidth]{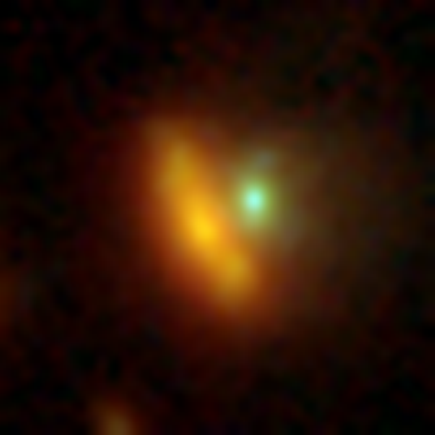}
    \includegraphics[width=0.19\textwidth]{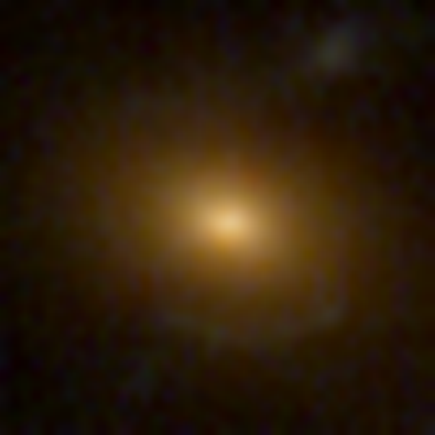}
    \\
    \includegraphics[width=0.19\textwidth]{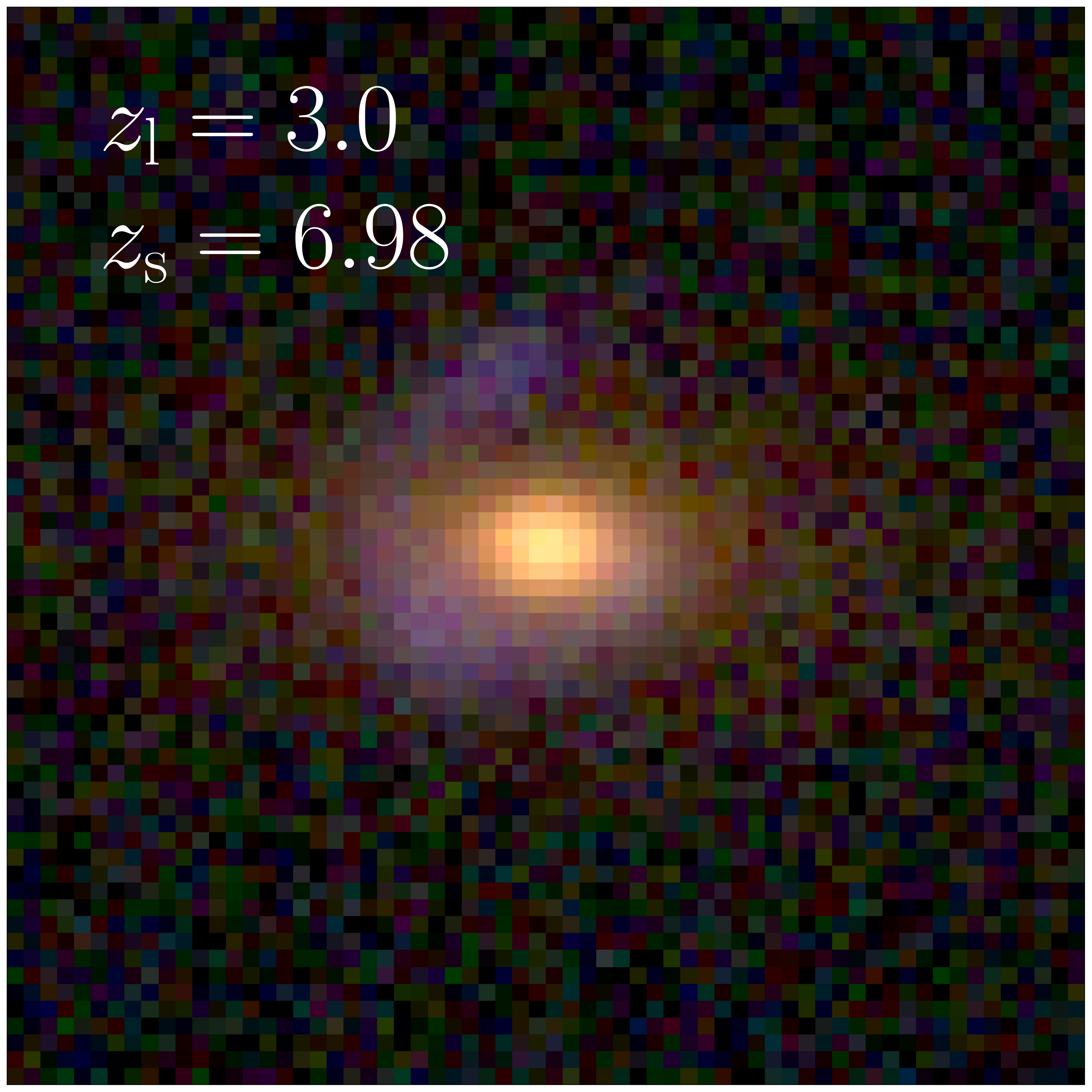}
    \includegraphics[width=0.19\textwidth]{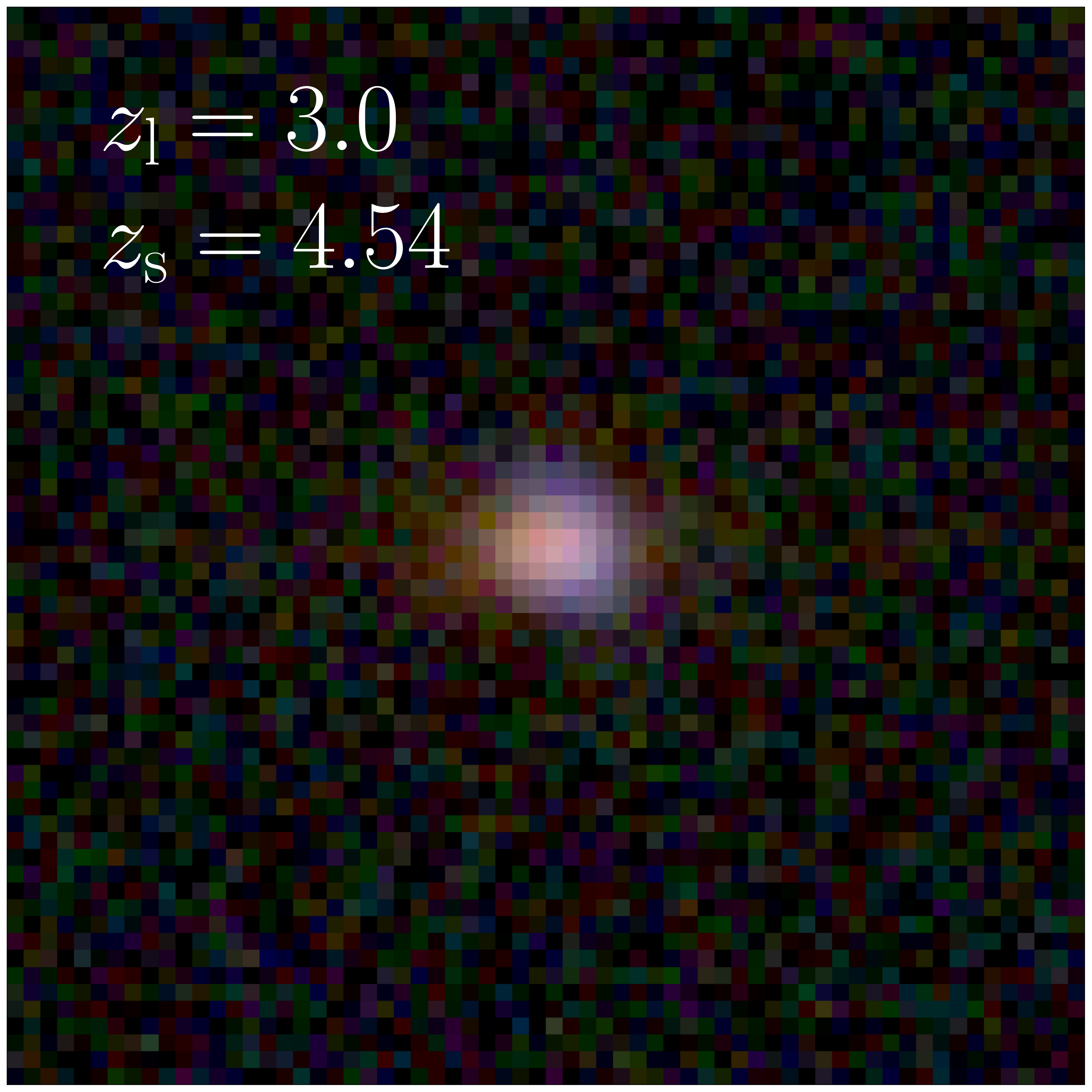}
    \includegraphics[width=0.19\textwidth]{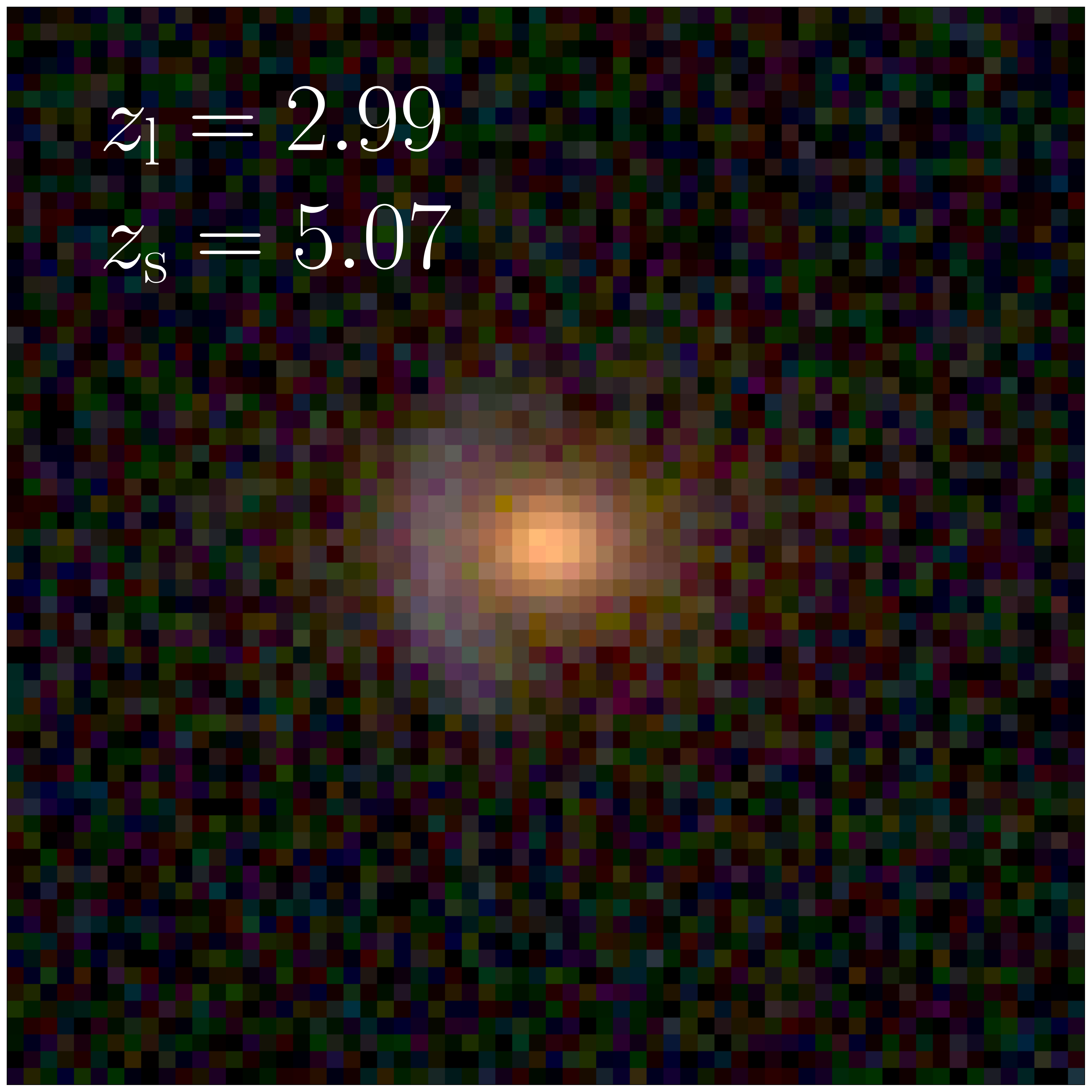}
    \includegraphics[width=0.19\textwidth]{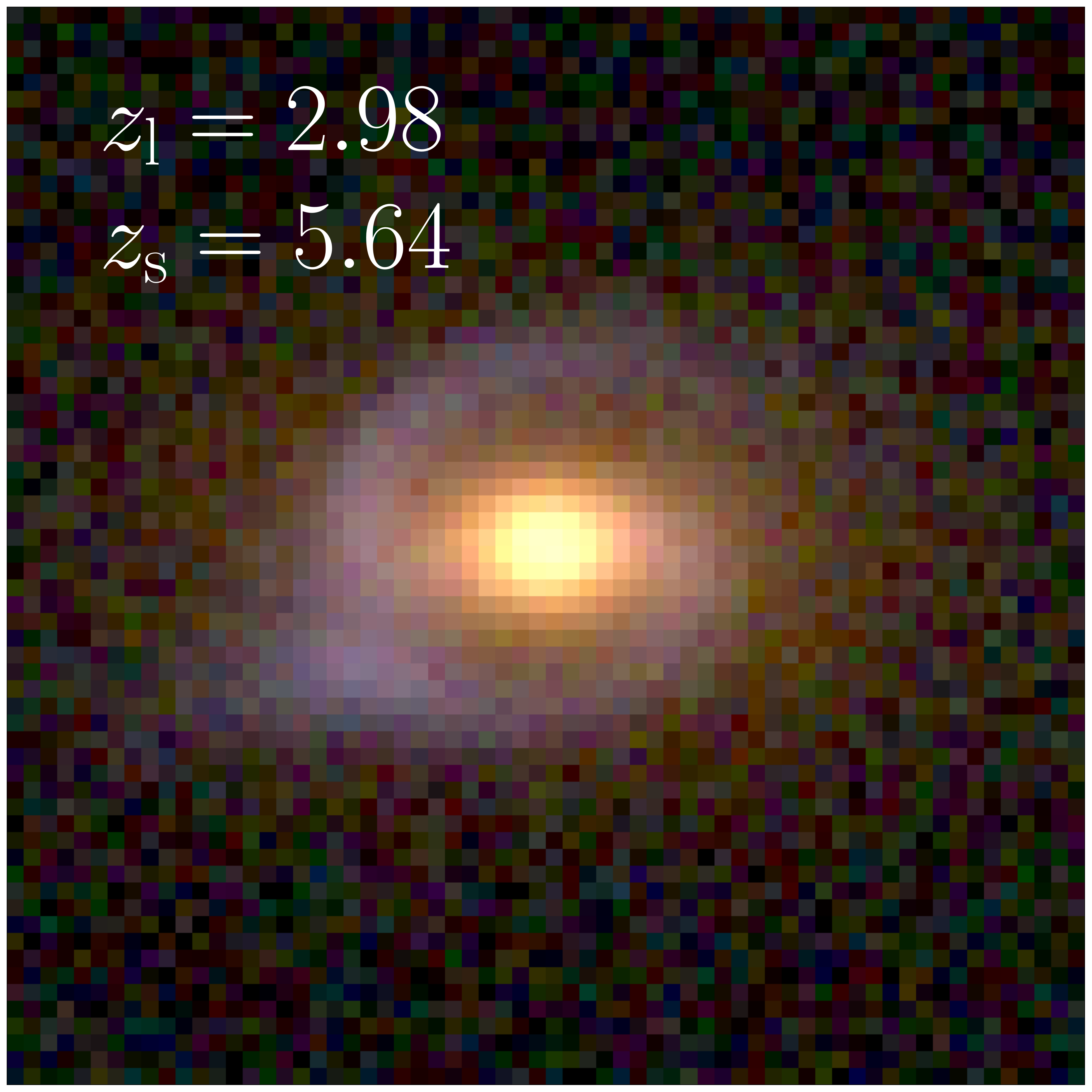}
    \includegraphics[width=0.19\textwidth]{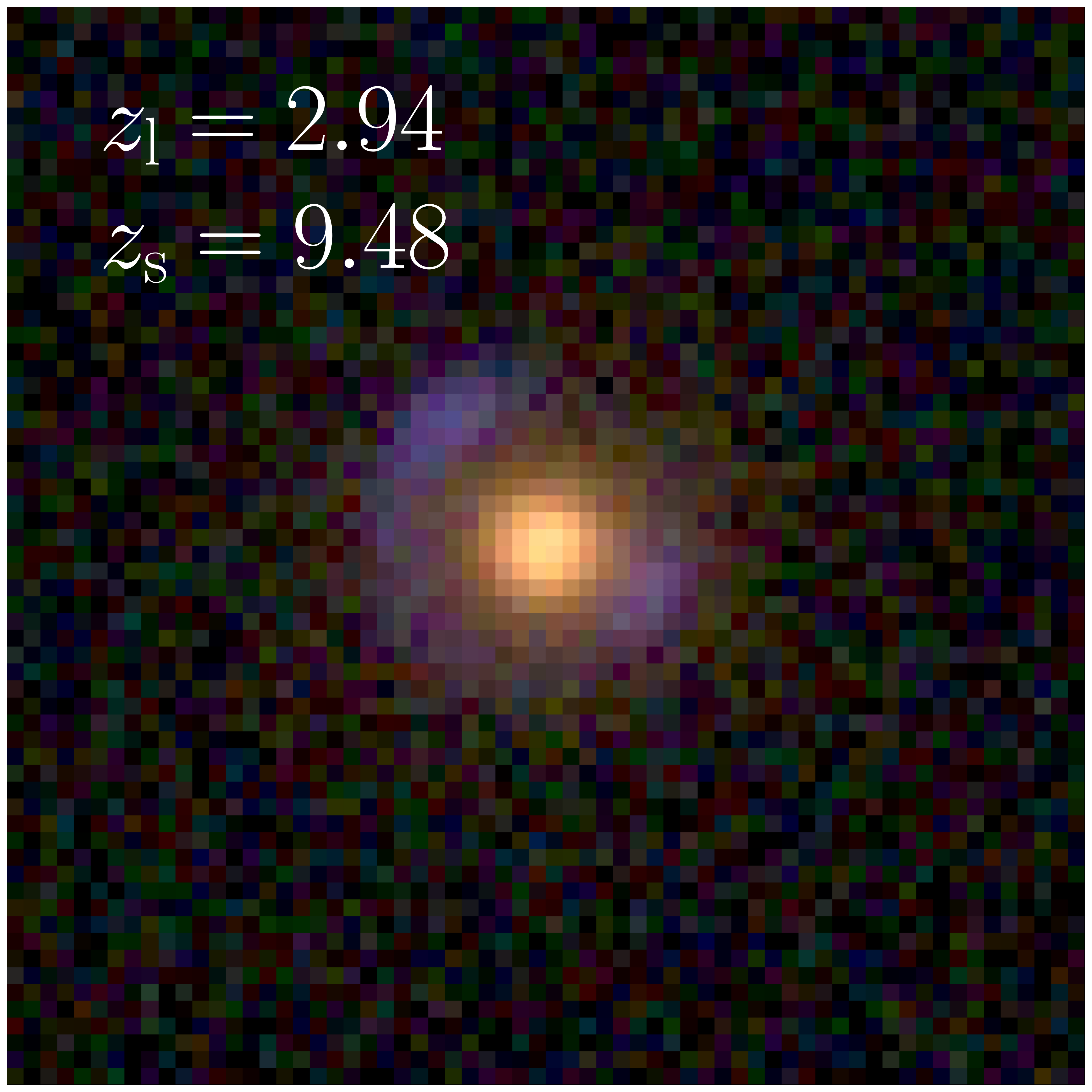}
    \caption{\textit{Top}: RGB images of the five systems with the highest photometric lens redshifts ($z_{\rm lens} > 1.8$) from the \Ntotalweak high and medium confidence lenses, excluding two of the spectacular lenses presented in \protect\cite{Mahler2025}. 
    \textit{Bottom}: a selection of five systems with the highest redshifts in our forecast. The lens and source redshifts are denoted in the top left of each panel. Colours are representative and are produced by combining the data simulated in three different NIRCam bands (F150W, F277W and F444W).}
    \label{fig:highz_obs_vs_sims}
\end{figure*}

To explore the question of high redshift lenses further, in the top row of \autoref{fig:highz_obs_vs_sims} we show the five systems with the highest photometric lens redshifts amongst the \Ntotalweak high and medium confidence lenses, excluding two of the spectacular lenses described in \cite{Mahler2025}. One of these two lenses is the COSMOS-Web ring, which has been described in detail in \cite{Mercier2024}, and at the time of discovery was the highest known lens redshift at $z=2.02$. The lenses and their associated photometric redshifts are listed in \autoref{tab:redshifts}. On the bottom row of the same figure, we show a selection of five simulated systems with the highest lens redshifts in our forecast.

% Einstein radii of the sims: 0.52, 0.14, 0.32, 0.61, 0.52

\begin{table}
    \centering
    \begin{tabular}{Slccc}
     Lens    & Score & $\theta_{\rm E}$ & $z_{\rm lens}$ \\
     \hline
     \hline
     COSJ100151+022347 & 6/12 & $0.33''$ & 3.58\\
     COSJ100120+022435 & 6/12 & $0.86''$ & 2.54\\
     COSJ100122+022132 & 9/12 & $0.52''$ & 2.49\\
     % CWeb Ring would be here, COSJ100024+015334 and z = 2.4539
     %COSJ100025+015245 & Spectacular & 2.4486\\
     COSJ100120+015834 & 7/12 & $0.62''$ & 2.44\\
     COSJ100134+022037 & 7/12 & $0.52''$ & 1.82\\
     \hline
     \\
    \end{tabular}
    \caption{Details of the five lenses with the highest photometric redshifts in the \catname catalogue, excluding two spectacular lenses presented in \protect\cite{Mahler2025}.}
    \label{tab:redshifts}
\end{table}

From this figure, we can see that, for the simulated lenses, the colours of the lens and source are readily distinguishable, and clear arc features are apparent in most of the systems despite their high redshift and consequent faintness. This lends weight to the idea that more strong lenses at $z > 2$ are identifiable in \textit{JWST} imaging. More variability is seen in the colours and morphology of the observed candidate lenses than in the simulated ones; this is due to the simplicity of the lens and source models employed in our \lenspop forecast. 

Furthermore, we can use the AD test to quantify the differences between the forecast and observed distributions. In this case, the null hypothesis of the two samples being drawn from the same underlying distribution is not rejected in the case of the \spectacularlenses spectacular lenses, but is rejected for both the \Ntotal high confidence and \Ntotalweak high and medium confidence lenses, at $95\%$ confidence. The full test results are shown in Appendix \ref{sec:stattables}.

The computation of the forecast lens redshifts relies purely on cosmology, with no astrophysics involved. This implies that the differences between the forecast and observed distributions stems from a selection effect in the observed data, rather than a fundamental physical difference in the creation of the two populations. For example, lenses at lower redshifts are likely to be brighter than those at higher redshifts, and therefore easier to identify during visual inspection. This may be a contributing factor in the discovery of more lenses at $z < 1$ than expected from the forecast. Confirmed spectroscopic redshifts are needed for all lenses before definitive conclusions can be drawn.

\subsubsection{Lens magnitudes}
\begin{figure*}
	\centering
	\includegraphics[width=0.98\textwidth]{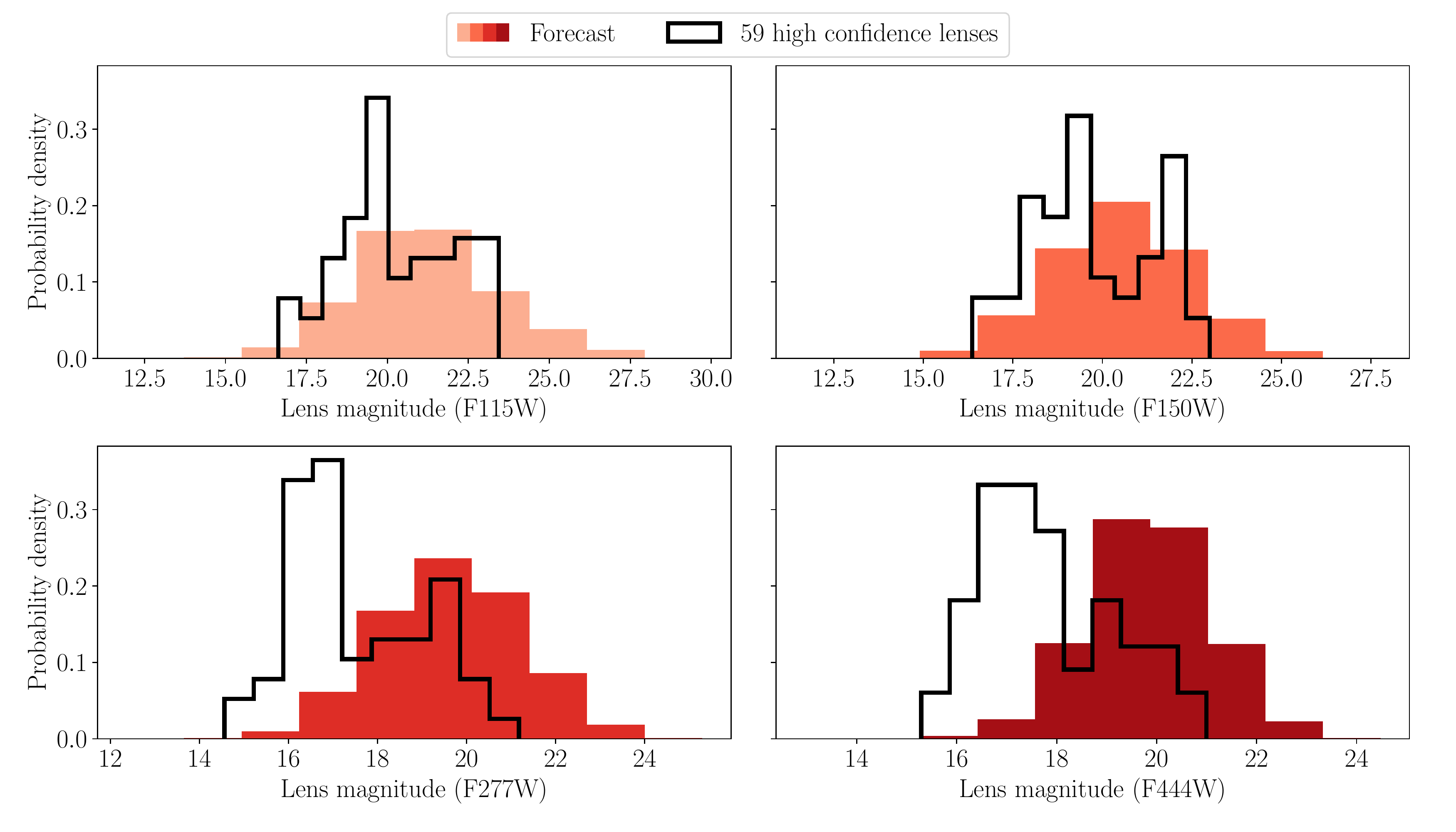}
	\caption{Our forecast distribution of lens magnitudes in each band (red) compared to the \Ntotal high confidence observed lenses (black).  }
	\label{fig:lensmagcomparison}
\end{figure*}

Next, we can compare the lens magnitudes in our forecast and observed populations of lenses. In \autoref{fig:lensmagcomparison}, we show the distribution of the AB magnitude of the lens galaxies in each NIRCam band in our forecast (red histograms) and in the \Ntotal high confidence lenses (unfilled black histograms). For brevity we choose to focus on the \Ntotal high confidence lenses, but the overall behaviour of the distributions in each band is the same for both the \spectacularlenses spectacular lenses and the \Ntotalweak high and medium confidence lenses.

From \autoref{fig:lensmagcomparison}, we can see that the distributions of the observed lens magnitudes tend to peak at brighter magnitudes than those of the forecast. Furthermore, the difference between the observed and forecast distributions increases with band wavelength. The AD test rejects the hypothesis that the two distributions were drawn from the same underlying distribution at 95\% confidence for all of the bands. 

There is also an indication of bimodality in the distribution of the observed lens magnitudes in each band.  To investigate this further, we performed a kernel density estimation to obtain the smoothed probability density function from the binned samples, which allowed us to identify that there is only a single mode in the distribution in each band \citep{Silverman1981}. We also carried out the dip test for deviation from a unimodal (uniform) distribution \citep{Hartigan1985}, finding no significant difference from a unimodal distribution for any of the bands. From this we conclude that any visual indication of bimodality is an artefact of the chosen binning scheme for the histograms and does not have an underlying physical cause. 

In general, we may conclude that the observed lenses are brighter in the F277W and F444W bands than would be expected from the \lenspop forecast. This finding has three possible underlying causes. Firstly, the estimation of the lens magnitudes by \cite{Nightingale2025} may have picked up a systematic bias during the lens modelling. However, if this were the case, we would expect all the bands to be affected equally, rather than what we actually see, which is that the effect is more pronounced in the long wavelength bands.

Secondly, if the \lenspop model is considered to be perfectly accurate, our finding could indicate that there remains an undiscovered population of faint lenses in the COSMOS-Web data. This is possible, since a magnitude cut was imposed before the beginning of the visual inspection, with only objects brighter than magnitude 23 in F277W being considered (which is clearly apparent in the distributions of the observed lens magnitudes in \autoref{fig:lensmagcomparison}). This possibility could easily be addressed by carrying out another visual inspection of the data with a more optimistic magnitude cut, for example examining all objects brighter than magnitude 26. 

Thirdly, we can consider the alternative possibility that the visual inspection is perfectly accurate, and that all lenses that exist in the COSMOS-Web data have been identified as such (this is already glossing over the truth, as we do not present or discuss any lens candidate that scored less than 6/12 in this work, an arbitrary but potentially consequential cut-off). In that case, it is the accuracy of the \lenspop model that must be examined.

Lens magnitudes in the \lenspop model are computed using the fundamental plane from \cite{Hyde:2008yf, Hyde:2008yh} derived from SDSS data. The fundamental plane is an empirically derived relation between galaxy properties \citep{Djorgovski1987, Dressler1987}, and can be expressed as $R \propto \sigma_{\rm v}^{\alpha}/I^{\beta}$, where $R$ is the effective radius of the light profile of the galaxy, $\sigma_{\rm v}$ is the velocity dispersion and $I$ is the surface brightness, and $\alpha$ and $\beta$ are parameters needing to be fit using a sample of observed galaxies. 

Again, considering that the visual inspection is perfectly accurate, our result suggests that magnitudes of lens galaxies observed in COSMOS-Web may not follow the same fundamental plane relation as that used in \lenspop, from galaxies catalogued in SDSS. This is in line with the fact that the \nht{spectroscopic redshifts from SDSS used to compute the fundamental plane do not extend above $z=0.35$ \citep{Hyde:2008yf}}, in stark contrast to our lens galaxy photometric redshift distribution which peaks at around $z=1$ and extends to above $z=3$. 

% distribution of galaxy photometric redshifts in SDSS does not extend above $z\approx 0.8$ \citep{Beck2016}, in stark contrast to our lens galaxy photometric redshift distribution which peaks at around $z=1$ and extends to above $z=3$. 

\nht{Whilst we cannot investigate this hypothesis in detail without spectroscopic redshifts and measurements of the velocity dispersions for the entire \catname catalogue, we can make a related comparison, between a stellar mass distribution computed from our forecast velocity dispersions, and the stellar masses of the observed lenses (obtained via photometry, see \cite{Nightingale2025}.}

\nht{\cite{Hyde:2008yf} provide the best-fit coefficients for linear and quadratic pairwise scaling relations between the various galaxy properties that are linked via the fundamental plane (Table 1 of that work). In particular, they provide linear and quadratic fits for the velocity dispersion as a function of stellar mass. We invert the quadratic relation to obtain stellar mass as a function of velocity dispersion. This allows us to compute a distribution of stellar masses from our forecast velocity dispersions, which can then be compared with the observed data. We show this comparison in \autoref{fig:mstarcomparison}.}

\begin{figure*}
	\centering
	\includegraphics[width=\textwidth]{plots/stellar_mass.pdf}
	\caption{Our forecast distribution of lens stellar mass (blue) compared to the observed lenses (black). \textit{Left}: the \spectacularlenses spectacular lenses. \textit{Middle}: the \Ntotal high confidence lenses. \textit{Right}: the \Ntotalweak high and medium confidence lenses.}
	\label{fig:mstarcomparison}
\end{figure*}

\nht{From this figure, we can see that the forecast and observed distributions of stellar mass match well for the 17 spectacular lenses, both peaking at around $M_* = 10^{11}M_\odot$. As we increase the sample size, we see that lenses with stellar masses well below that predicted by our forecast begin to appear, with a very large low-mass tail in the distribution for the high and medium confidence lenses.}

\nht{One interpretation of this finding is that the velocity dispersion function produced by \cite{Choi2007} cannot correctly predict lens stellar mass at high redshifts, resulting in our forecast predicting too many massive galaxies at high redshifts, $z>2$. A simple check of this hypothesis is to plot our forecast and observed lens redshifts against stellar mass, to see if the correlations are similar across the two populations.}

\begin{figure}
	\centering
	\includegraphics[width=0.47\textwidth]{plots/zlens_vs_mstar.pdf}
	\caption{A scatter plot of lens redshift versus lens stellar mass for our forecast population (blue) and 59 high confidence observed lenses (black).}
	\label{fig:zlensmstar}
\end{figure}

\nht{In \autoref{fig:zlensmstar}, we show a scatter plot of lens redshift against lens stellar mass for our forecast and observed populations, showing the 59 high confidence lenses as a representative example. We also plot the lines of best fit to these data computed via linear regression. From this figure, we can see that, at $z< 2$, i.e. at \textit{low} redshifts, our forecast contains more massive galaxies than seen in the data. Conversely, as redshift increases, the two populations appear to be in better agreement. However, since there are so few observed lenses at $z>2$, it is not appropriate to draw strong conclusions about the true agreement of the two populations.}

\nht{It was also pointed out in COWLS Paper I \citep{Nightingale2025}, that the lens candidate score and lens stellar mass are correlated, with higher scoring candidates tending to be the most massive lenses. If we \textit{do} trust the extrapolation of the \cite{Choi2007} velocity dispersion function to $z>2$, our finding may hint that the low-scoring, low-mass candidates are not true strong lenses, as they are not present in the forecast. We reiterate that this comparison itself relies on the fundamental plane relations fit by \cite{Hyde:2008yf}. A more robust test of how well various velocity dispersion functions match the data will only be possible with spectroscopy for our lens candidates.}

\subsubsection{Unlensed source magnitudes}
\begin{figure*}
	\centering
	\includegraphics[width=0.98\textwidth]{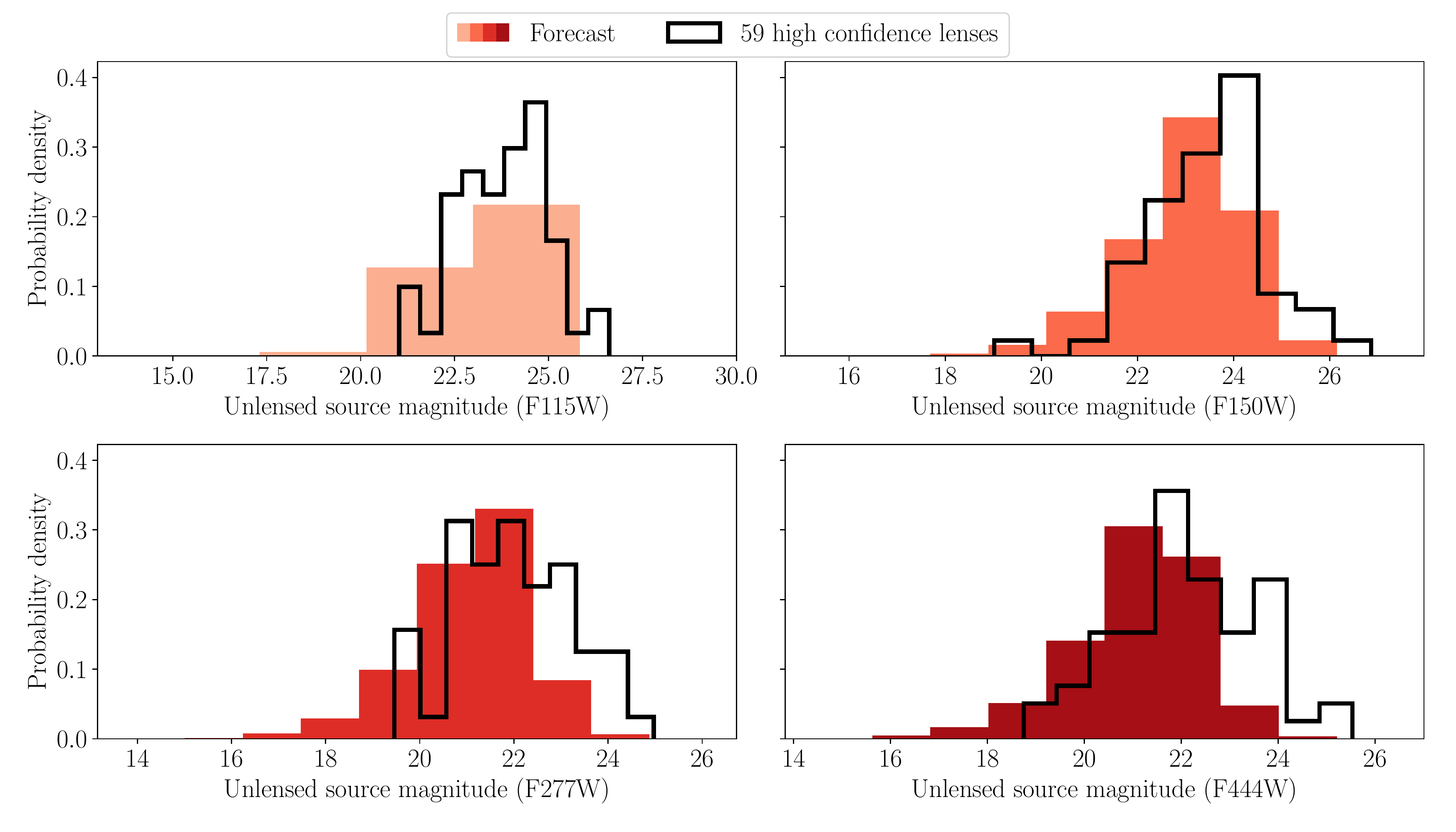}
	\caption{Our forecast distribution of unlensed source magnitudes in each band (red) compared to the \Ntotal high confidence observed lenses (black).}
	\label{fig:sourcemagcomparison}
\end{figure*}

Lastly, we can compare the unlensed source magnitudes in our forecast and observed populations of lenses, i.e. the magnitudes of the sources in the source planes. In \autoref{fig:sourcemagcomparison}, we show the distributions of this quantity in each of the NIRCam bands for the forecast lenses (red histograms) and the \Ntotal high confidence observed lenses (black unfilled histograms). As with the lens magnitudes, we choose this set of lenses as a representative example. From this figure, we can see that we have relatively good agreement in the overall shape and peak location of the source magnitude distribution in each band. However, the AD test rejects the null hypothesis of similarity between the forecast and observed distributions. In particular, we can see that, as the wavelength increases, the observed lenses contain an over-abundance of faint sources and a lack of bright sources compared to the forecast. 

However, the forecast source magnitudes are not computed by \lenspop, but are instead taken directly from the JAGUAR source catalogue. The explanation of the difference between the observed and forecast distributions therefore again has more than one possible explanation. Firstly, the magnitudes in the JAGUAR catalogue may be inaccurate, but this is unlikely as the mock galaxies in the catalogue were produced from an empirical model based solely on observational data \citep{Williams2018}. One potentially relevant factor in the production of the JAGUAR mock is the use of the \cite{Bruzual:2003tq} stellar population synthesis model via \texttt{BEAGLE}; differences in such models have been shown to propagate to the galaxy SEDs \citep{Newman2025}. Nevertheless, unless the \catname strong lens sources are a particularly unusual sample of galaxies, we can expect their properties to match well with the galaxies in the JAGUAR mock catalogue. 

Secondly, the computation of the unlensed source magnitudes for the observed lenses may carry some inaccuracies. The observed lensed sources were reconstructed by \cite{Nightingale2025} using a multi-Gaussian expansion and Voronoi mesh approach~\citep{He:2024udi}. The magnitudes are computed from the resulting reconstructed source, and will naturally depend on the inferred lens model and final source mesh configuration. How much the modelling choices affect the source reconstruction cannot be examined in detail until different lens and source models have been applied to the COWLS catalogue.

A final contribution to the differences between the forecast and observed distributions may be a selection effect, but we would generally expect the visual inspection to select for preferentially brighter sources rather than fainter ones, as bright images are easier for human visual inspectors to identify and examine than faint ones. Since we see almost the opposite effect here, this may lead us to conclude that some bright sources have been missed in the visual inspection, whilst some of the (likely weaker) candidate lenses with faint sources may be false positives.

\subsection{Comparison with other forecasts for COSMOS-Web}

\subsubsection{Holloway et al. (2023)}
\cite{Holloway:2023axl} performed a forecast for a number of \textit{JWST} surveys, including COSMOS-Web. Instead of relying on simulated lenses, they used mock galaxies from the JAGUAR catalogue as both sources and lenses. Furthermore, they modified the effective radii of the deflectors to better match the S\'ersic indices in the catalogue, and assigned halo masses from the DREaM catalogue \citep{Drakos:2021bsb} to the JAGUAR galaxies using the subhalo abundance matching technique. This work predicted $\sim 65$ strong lenses across all bands in COSMOS-Web, compared to the 107 we forecast using \lenspop and the \Ntotalweak high and medium confidence lenses in the \catname catalogue from the observed COSMOS-Web data. 

In order to investigate the source of the difference between the forecasts, we firstly re-ran our \lenspop analysis using the modified JAGUAR catalogue produced by \cite{Holloway:2023axl} as the source population. This resulted in the same number of strong lenses predicted as well as distributions for the lens system properties consistent with our original findings, meaning that the difference between the results of this work and that of \cite{Holloway:2023axl} is not driven by the differences in the source catalogues. However, the \cite{Holloway:2023axl} prediction of $3.1\times 10^9$ strong lenses in the Universe is larger than our finding of $2.8\times 10^8$ lenses. Combined with the smaller number of strong lenses predicted by \cite{Holloway:2023axl}, this implies that the difference in the number of detectable lenses in COSMOS-Web is likely the result of differences in how the survey specifications were applied to the idealised lens population. 

Examining the properties of the forecast lens populations, we find that $<1\%$ of our forecast lenses are fainter than the magnitude cut applied by \cite{Holloway:2023axl} in the F115W band, with no lenses in any of the other three bands being fainter than their respective cuts. The smallest Einstein radius of a lens in the \cite{Holloway:2023axl} forecast lens catalogue is 0.07 arcseconds, whilst our forecast predicts $0.01\%$ of lenses having Einstein radii between 0.04 and 0.07 arcseconds. The source and lens redshift distributions predicted by \cite{Holloway:2023axl} are very similar to those produced by \lenspop. The discrepancy between the predictions therefore cannot be explained by \lenspop predicting more fainter or smaller lenses in the population than the method of \cite{Holloway:2023axl}. 

Lastly, we note that \cite{Holloway:2023axl} attempted to validate their forecast by comparing their predictions for lensing abundance in the COSMOS field with previous observations of the field made using \textit{HST}. In particular, \cite{Faure:2008dt} found 67 candidate strong lenses, compared to the prediction by \cite{Holloway:2023axl} of 34. \cite{Jackson:2008my} identified two certain, one highly probable and 112 further possible but unlikely lens systems, leading to $160$ candidates in total; applying the same search criteria, \cite{Holloway:2023axl} found 80. Taken together, these results could indicate a potential systematic underestimation of strong lensing abundance in their forecasts. This may be supported by the discovery of \Ntotalweak high and medium confidence strong lenses in COSMOS-Web, but if only the \Ntotal high confidence strong lenses are considered, the forecast of \cite{Holloway:2023axl} is in excellent agreement with the data.

\subsubsection{Ferrami \& Wyithe (2024)}
\cite{Ferrami:2024obm} built a model for strong lensing statistics which is more akin to the \lenspop method, using the assumption of all deflectors being SIEs to calculate the strong lensing optical depth. However, instead of relying on a mock galaxy catalogue to simulate the source population, they computed the number density of sources lensed by a given deflector analytically. 

Their work tested the effect of different velocity dispersion functions on the number of lenses detectable by \textit{Euclid}, with their fiducial model being that of \cite{Mason:2015wla}, which is redshift dependent. They compared properties of their forecast sample with one produced using the \cite{Choi2007} velocity dispersion function employed in this work (\autoref{eq:veldisp}). They found that the fiducial \cite{Mason:2015wla} model results in deflector and source redshift distributions which peak at a slightly lower redshift compared to the \cite{Choi2007} model, and a distribution of Einstein radii which peaks at a higher value compared to the \cite{Choi2007} model.

\cite{Ferrami:2024obm} did not produce a forecast for COSMOS-Web using the \cite{Choi2007} velocity dispersion function, but using their fiducial \cite{Mason:2015wla} model they predicted 46 lenses detectable in the F115W filter, 74 in the F150W filter and 121 in the F277W filter, all assuming lens light subtraction. These numbers are broadly consistent with our prediction of 107 lenses across all bands. They did not produce a forecast for the F444W filter.

However, the forecast distributions of deflector and source redshifts, velocity dispersion and Einstein radii in each band do not match those predicted with our \lenspop model. In particular, the source redshift distribution is consistent with zero at $z\sim 5$, unlike our finding of source redshifts up to $z\sim 11$, likely due to our use of the JAGUAR catalogue which contains sources up to $z\sim15$. 

Furthermore, the tail of our distribution of Einstein radii reaches about 2 arcseconds, compared to around 3 arcseconds in \cite{Ferrami:2024obm}. As discussed in \autoref{subsec:properties}, these differences are correlated, as Einstein radii are larger for lenses at lower redshifts for a given source redshift. Given the similarity in the forecasting methods, the cause of these differences is likely the different velocity dispersion function used by \cite{Ferrami:2024obm}, which results in their forecast containing a larger number of massive galaxies than ours. 

%\cite{Collett2015} investigated the effect of the \cite{Mason:2015wla} model on a forecast for the abundance of lenses in DES, finding an increase compared to the \cite{Choi2007} model, but with no comment about any change in parameter distributions.

\subsection{Selection effects}\label{subsec:selection}
All of the comparisons we have made so far are founded on the assumption that the observed and forecast lenses were selected in exactly the same way. Whilst the criteria of \cite{Collett:2015roa}, which we also employ (see \autoref{subsec:lenses}), are designed to select for lenses in the same way that visual inspectors decide whether an object exhibits lensing features, the correspondence is not one to one. 

For example, the first criterion that a candidate must meet to be deemed a lens in our forecast is that multiple images of the source must be produced. This criterion is baked in to the calculation of the strong lensing optical depth. But in the second round of visual inspection described in \cite{Nightingale2025}, inspectors had the option to grade a lens as `S', meaning that only a singly imaged lensing feature was visible\footnote{We note that in many cases, the presence of the counter-image was revealed by the lens modelling conducted in that work.}. An `S' grade contributed one point towards the total score of a candidate; 47 candidates in the complete COWLS catalogue received at least one `S' grade (11.2\%).

\nht{A full forward modelling of the selection function to understand the bias of our lenses with respect to the parent galaxy population goes beyond the scope of this paper}. Such efforts have been made for other lens populations, see e.g. \cite{Sonnenfeld2022, Sonnenfeld2024}. We here describe the various contributing factors to the selection of the forecast and observed lenses and provide a brief discussion of how they may have affected the results described in this work.

% \nhc{check carefully with James' paper that the discussion here is not too repetitive}

\subsubsection{Selection of the forecast lenses}
As a reminder, the criteria we use to call a forecast lens detectable are that multiple images are produced, the image and counter-image must be resolved, and the total source magnification and signal-to-noise ratio must exceed a given threshold. To investigate the effects of these choices on the detectable lens population, we can modify the chosen thresholds.

% this stuff used to be at the end of section 4
%Lastly, we tested the effect that changing the detectability criteria described in \autoref{subsec:lenses} has on the forecast strong lens population. 
Keeping the other criteria fixed, we firstly increased the signal-to-noise ratio required to call a lens detectable from 20 to 40; consequently, the number of detectable lenses decreased to 55, almost exactly half the original number of 107. The distribution of the Einstein radii of these lenses does not contain the very smallest Einstein radii lenses of our original forecast, as expected from the fact that lenses with smaller Einstein radii will generally have lower signal-to-noise ratios. Furthermore, the distribution of velocity dispersions does not contain the lowest mass galaxies, and the source redshift distribution misses the vast majority of sources above $z=6$; the unlensed source magnitude distributions miss the faintest sources. In contrast, the distribution of lens redshifts and lens magnitudes remain consistent with our original result.

Increasing the threshold for the minimum source magnification from 3 to 4 results in a forecast of 68 detectable lenses, with the distributions of all the parameters remaining close to the original results. Increasing the resolution threshold at which a lens is determined to be detectable, from $\theta_{\rm E}^2 > r_{\rm s}^2 + (s/2)^2$ to $\theta_{\rm E}^2 > r_{\rm s}^2 + s^2$, results in 104 lenses, with distributions again consistent with the original results. This indicates that the signal-to-noise ratio threshold is by far the most important criterion for determining strong lensing detectability, excluding the requirement that multiple images of the source must be produced.

\subsubsection{Selection of the observed lenses}
The first and most obvious explicit selection the COWLS catalogue was subject to is the cut in magnitude that was imposed to choose objects for visual inspection. Only those objects brighter than AB magnitude 23 in the F277W filter were inspected; whilst our forecast predicts that $1.5\%$ of the COWLS lenses should lie above this magnitude, by construction we find none. Furthermore, as we have seen, our forecast predicts significantly more fainter lenses in general than what is found in COWLS: 41\% of the forecast lenses are fainter than magnitude 20 in the F277W in the forecast, compared to just $5\%$ in COWLS.

Another important selection made in the preparation of this work was the choice to only consider those COWLS candidates with scores of 6/12 or more. The forecast, by construction, is completely comprised of true lenses which should all score 12/12 in an inspection; limiting our comparison with the forecast to only those COWLS candidates that scored 12/12 (all \spectacularlenses spectacular lenses from \cite{Mahler2025} plus two lenses from \citealt{Nightingale2025}) would not have been justified from a statistical standpoint. A way to quantify the selection effect of the grading scheme would be to have visual inspectors examine simulated images of the forecast lenses, like those shown in \autoref{fig:long_grid} and \autoref{fig:highz_obs_vs_sims}, to see how many ostensibly 12/12 lenses are actually graded as such. This would still not be a perfect test, as the SIE lens model in the forecast tends to result in regular, unambiguous lensing features. 

The final, and implicit, selection effect is that of human fallibility. The first half of the first round of visual inspection in COWLS was conducted by four inspectors, and the second half by five; the second round was conducted by six inspectors. When analysing the ability of a visual inspection team to correctly identify lenses, \cite{DES:2023hxm} found that the minimum number of inspectors needed to mitigate noise is six. Future visual inspections of the COSMOS-Web field for strong lenses could be aided by increasing the number of inspectors or by using machine-learning search methods.

\section{Conclusions} \label{sec:conclusions}
In this work we made a forecast for the abundance and properties of strong gravitational lenses detectable in the COSMOS-Web survey, a survey of the 0.54 deg$^2$ COSMOS field using the NIRCam and MIRI instruments aboard \textit{JWST}. We used the \lenspop model of \cite{Collett:2015roa} to make our forecast, after implementing several modifications to improve its suitability as a forecasting tool for COSMOS-Web. In particular, we changed the mock galaxy catalogue from which the source population is drawn to the JAGUAR catalogue, and added the NIRCam filter data and COSMOS-Web survey specifications. This led to a prediction of 107 lenses detectable across all bands, or a minimum of 14 detectable in the F115W$-$F150W, F115W$-$F277W and F115W$-$F444W difference images. 

Our forecast predicted the strong lenses in COSMOS-Web are dominated by small Einstein radius systems ($\theta_{\rm E} < 1^{\prime \prime}$) with a distribution of velocity dispersions that peaks at $\sigma_{\rm v} = 200$ kms$^{-1}$. The distribution of lens redshifts is predicted to peak at $z=1$, with a significant tail up to $z=3$. The distribution of source redshifts peaks at $z=3.2$, with a non-negligible fraction of the sources lying at $z>6$, directly probing the entire epoch of reionisation. The highest source redshift in our forecast population is $z=11.6$.

We then compared the properties of the forecast strong lenses with that of the observed data from the \catname catalogue. Strong lenses were identified in COSMOS-Web by two rounds of visual inspection; the first round found \spectacularlenses strong lenses whilst the second round which included modelling of 419 additional candidates found \halflenses lens candidates that scored at least 6/12; there were \threequarterlenses lens candidates that scored at least 8/12. This yielded a total catalogue of \Ntotal high confidence strong lenses, or \Ntotalweak high and medium confidence strong lenses.

The distribution of Einstein radii in the \spectacularlenses spectacular lenses does not match that of our \lenspop forecast, with both the KS and AD tests rejecting the null hypothesis that these samples were drawn from the same underlying distribution at $95\%$ confidence. This is driven by a lack of lenses with small Einstein radii in the observed sample compared to the forecast. However, the null hypothesis is not rejected by these tests when comparing the distribution of Einstein radii for the \Ntotal high confidence lenses or the \Ntotalweak high and medium confidence lenses.

In contrast, the distributions of the lens photometric redshifts, lens magnitudes and unlensed source magnitudes are not consistent with the forecast distributions for these quantities. In particular, there are fewer lenses above $z=1$ in the observed sample than in the forecast, and the lens magnitude distributions peak at brighter magnitudes than expected from the forecast. This may be due to the computation relying on a fundamental plane relation fit with SDSS data in which the galaxies are at much lower redshifts than the typical \catname lens galaxy. 

The unlensed source magnitudes appear to be slightly fainter in the observed sample than would be expected from the forecast, particularly in the long wavelength bands; the use of the JAGUAR mock catalogue means that the forecast should be a good representation of galaxies observable in COSMOS-Web, even up to high redshifts. Modelling the COWLS sample with a variety of lens and source models may yield better agreement in the source magnitudes than is currently seen. A more in-depth comparison of the lens and source galaxies with the \lenspop forecast will only be possible once velocity dispersion and spectroscopic redshift information are obtained, which will require follow-up observations from \textit{JWST} to obtain the required precision at the redshifts in question.

% in respect to the above: I used the shortwavelength pixel size (0.031) for the entire forecast, because I would have had to hack the code (a lot?) to pass two separate pixel sizes. could this account for the bigger differences in the F227W and F444W magnitudes between forecast and data compared to the other bands? hmm

It is also important to note that the \lenspop model has a number of simplifications: firstly, the images are convolved with a Gaussian PSF rather than the more complex PSF of \textit{JWST}; secondly, that all the lenses are modelled as SIEs, meaning that the images produced are typically simple rings, arcs or multiple images, unlike realistic lenses; and lastly that the weak lensing by line of sight objects (i.e. ``external shear'') is not taken into account. However, it is unlikely that these assumptions significantly impact the forecast abundance of strong lenses. As described in \autoref{subsec:selection}, the signal-to-noise ratio plays the largest role in the detectability, and signal-to-noise ratio is not strongly correlated with the complexity of lensed image morphology. 
% I think this is true

Lastly, we reiterate that the MIRI observations of COSMOS-Web have not yet been inspected for strong lenses. Made using the single F770W filter, the MIRI observations cover a smaller sky area, $0.19$~deg$^2$, than those using NIRCam, but we may expect another handful of lenses to be detected in these data. This is because MIRI observes even longer wavelengths than NIRCam, which will reveal even redder and potentially higher redshift systems than those in the current COWLS catalogue, akin to how NIRCam has revealed very red lensed sources which are invisible in \textit{HST} imaging \citep{Mahler2025}. The MIRI-bright lenses are likely to have Einstein radii comparatively larger than the majority identified in the NIRCam data, due to the lower spatial resolution of MIRI, which means that smaller Einstein radius lenses are less likely to be resolved. 

In conclusion, we have shown that the \lenspop strong lens forecast model can be successfully adapted for \textit{JWST} observations, and that the abundance of strong lenses and the distribution of the Einstein radii of observed lenses in COSMOS-Web are in good agreement with \lenspop predictions. This implies that our choice of source galaxy population, the JAGUAR mock catalogue, is a good match to strong lens sources from the current day to high redshifts, $0 \leq z \leq 11$, spanning $92\%$ of the Universe's history. Furthermore, it supports the idea that the majority of strong lenses can be described by elliptical isothermal mass profiles. 

However, the distributions of the magnitudes of the observed deflectors and sources do not align as well with those from our forecast, with lenses in the observed data being systematically brighter than expected from the forecast. This result motivates detailed follow-up of the entire \catname catalogue to obtain spectroscopic lens and source redshifts and lens velocity dispersions, in order to better understand the source of these discrepancies, as well as targeted re-inspection of the COSMOS-Web field to search for the undiscovered population of faint, high-redshift lenses. 

% In conclusion, we have shown that the \lenspop strong lens forecast model can be adapted for \textit{JWST} observations, and that the abundance of strong lenses and the distribution of the Einstein radii of observed lenses in COSMOS-Web are in good agreement with \lenspop predictions. However, the distributions of the magnitudes of the deflectors and sources do not align with those from our forecast. Our results motivate detailed follow-up of the entire \catname catalogue to obtain lens and source redshifts and lens velocity dispersions, in order to better understand the source of these discrepancies. 

% say something abt archival data here? i.e. everyone should go and look for lenses in their data?

\section*{Acknowledgements}
We are grateful to Tom Collett for useful discussions about \lenspop, Giovanni Ferrami for discussions about \cite{Ferrami:2024obm}, and to Philip Holloway for providing us with the modified version of the JAGUAR catalogue created for \cite{Holloway:2023axl} along with his catalogues of forecast lens properties. \nht{We thank the referee for their detailed comments which helped improve the paper.}
%We thank Kaija Virolainen for contributing to the production of the RGB images shown in \autoref{fig:highz_obs_vs_sims}. 

NBH is supported by a postdoctoral position funded by IN2P3. JWN is supported by an STFC/UKRI Ernest Rutherford Fellowship, Project Reference: ST/X003086/1. AA and QH acknowledge support from the European Research Council (ERC) through Advanced Investigator grant DMIDAS (GA 786910). GM and RM were supported in Durham by STFC via grant ST/X001075/1, and the UK Space Agency via grant ST/X001997/1. DS acknowledges the Jet Propulsion Laboratory, California Institute of Technology, under a contract with the National Aeronautics and Space Administration (80NM0018D0004). OI acknowledges the funding of the French Agence Nationale de la Recherche for the project iMAGE (grant ANR-22-CE31-0007).

This research has made use of the SVO Filter Profile Service ``Carlos Rodrigo'', funded by MCIN/AEI/10.13039/501100011033/ through grant PID2023-146210NB-I00. This work was made possible by utilising the CANDIDE cluster at the Institut d’Astrophysique de Paris. The cluster was funded through grants from the PNCG, CNES, DIM-ACAV, the \textit{Euclid} Consortium, and the Danish National Research Foundation Cosmic Dawn Center (DNRF140). It is maintained by Stephane Rouberol.  The French contingent of the COSMOS team is partly supported by the Centre National d'\'Etudes Spatiales (CNES).

This work is based on observations made with the NASA/ESA/CSA James Webb Space Telescope. The data were obtained from the Mikulski Archive for Space Telescopes at the Space Telescope Science Institute, which is operated by the Association of Universities for Research in Astronomy, Inc., under NASA contract NAS 5-03127 for JWST. These observations are associated with program 1727.

\section*{Data Availability}

The forecast data underlying this article were generated with the publicly available modified version of \lenspop~\href{https://github.com/nataliehogg/lenspop}{\faGithub}; the forecast lens catalogue is available on request. The observational data are publicly available in the COWLS repository~\href{https://github.com/Jammy2211/COWLS_COSMOS_Web_Lens_Survey}{\faGithub}.

\bibliographystyle{mnras}
\bibliography{cweb_forecast}

\begin{thebibliography}{}
\makeatletter
\relax
\def\mn@urlcharsother{\let\do\@makeother \do\$\do\&\do\#\do\^\do\_\do\%\do\~}
\def\mn@doi{\begingroup\mn@urlcharsother \@ifnextchar [ {\mn@doi@} {\mn@doi@[]}}
\def\mn@doi@[#1]#2{\def\@tempa{#1}\ifx\@tempa\@empty \href {http://dx.doi.org/#2} {doi:#2}\else \href {http://dx.doi.org/#2} {#1}\fi \endgroup}
\def\mn@eprint#1#2{\mn@eprint@#1:#2::\@nil}
\def\mn@eprint@arXiv#1{\href {http://arxiv.org/abs/#1} {{\tt arXiv:#1}}}
\def\mn@eprint@dblp#1{\href {http://dblp.uni-trier.de/rec/bibtex/#1.xml} {dblp:#1}}
\def\mn@eprint@#1:#2:#3:#4\@nil{\def\@tempa {#1}\def\@tempb {#2}\def\@tempc {#3}\ifx \@tempc \@empty \let \@tempc \@tempb \let \@tempb \@tempa \fi \ifx \@tempb \@empty \def\@tempb {arXiv}\fi \@ifundefined {mn@eprint@\@tempb}{\@tempb:\@tempc}{\expandafter \expandafter \csname mn@eprint@\@tempb\endcsname \expandafter{\@tempc}}}

\bibitem[\protect\citeauthoryear{{Abbott} et~al.}{{Abbott} et~al.}{2018}]{DESDR1}
{Abbott} T.~M.~C.,  et~al., 2018, \mn@doi [The Astrophysical Journal Supplement] {10.3847/1538-4365/aae9f0}, \href {https://ui.adsabs.harvard.edu/abs/2018ApJS..239...18A} {239, 18}

\bibitem[\protect\citeauthoryear{Acevedo~Barroso et~al.}{Acevedo~Barroso et~al.}{2024}]{Euclid:2024jyk}
Acevedo~Barroso J.~A.,  et~al., 2024, {Euclid: The Early Release Observations Lens Search Experiment} (\mn@eprint {arXiv} {2408.06217})

\bibitem[\protect\citeauthoryear{Adelman-McCarthy et~al.}{Adelman-McCarthy et~al.}{2007}]{SDSS:2007aih}
Adelman-McCarthy J.~K.,  et~al., 2007, \mn@doi [The Astrophysical Journal Supplement] {10.1086/518864}, 172, 634

\bibitem[\protect\citeauthoryear{{Aihara} et~al.}{{Aihara} et~al.}{2018}]{HSCDR1}
{Aihara} H.,  et~al., 2018, \mn@doi [Publications of the Astronomical Society of Japan] {10.1093/pasj/psx081}, \href {https://ui.adsabs.harvard.edu/abs/2018PASJ...70S...8A} {70, S8}

\bibitem[\protect\citeauthoryear{Anderson \& Darling}{Anderson \& Darling}{1952}]{AndersonDarling1952}
Anderson T.~W.,  Darling D.~A.,  1952, \mn@doi [The Annals of Mathematical Statistics] {10.1214/aoms/1177729437}, 23, 193

\bibitem[\protect\citeauthoryear{{Asada} et~al.,}{{Asada} et~al.}{2023}]{Asada2023}
{Asada} Y.,  et~al., 2023, \mn@doi [Monthly Notices of the Royal Astronomical Society] {10.1093/mnrasl/slad054}, \href {https://ui.adsabs.harvard.edu/abs/2023MNRAS.523L..40A} {523, L40}

\bibitem[\protect\citeauthoryear{{Beckwith} et~al.,}{{Beckwith} et~al.}{2006}]{Beckwith2006}
{Beckwith} S. V.~W.,  et~al., 2006, \mn@doi [The Astronomical Journal] {10.1086/507302}, \href {https://ui.adsabs.harvard.edu/abs/2006AJ....132.1729B} {132, 1729}

\bibitem[\protect\citeauthoryear{Birrer \& Amara}{Birrer \& Amara}{2018}]{Birrer:2018xgm}
Birrer S.,  Amara A.,  2018, \mn@doi [Physics of the Dark Universe] {https://doi.org/10.1016/j.dark.2018.11.002}, 22, 189

\bibitem[\protect\citeauthoryear{Birrer, Welschen, Amara  \& Refregier}{Birrer et~al.}{2017}]{Birrer:2016xku}
Birrer S.,  Welschen C.,  Amara A.,   Refregier A.,  2017, \mn@doi [Journal of Cosmology and Astroparticle Physics] {10.1088/1475-7516/2017/04/049}, 04, 049

\bibitem[\protect\citeauthoryear{Birrer, Refregier  \& Amara}{Birrer et~al.}{2018}]{Birrer:2017sge}
Birrer S.,  Refregier A.,   Amara A.,  2018, \mn@doi [The Astrophysical Journal] {10.3847/2041-8213/aaa1de}, 852, L14

\bibitem[\protect\citeauthoryear{Birrer et~al.,}{Birrer et~al.}{2021}]{Birrer2021}
Birrer S.,  et~al., 2021, \mn@doi [Journal of Open Source Software] {10.21105/joss.03283}, 6, 3283

\bibitem[\protect\citeauthoryear{Bruzual \& Charlot}{Bruzual \& Charlot}{2003}]{Bruzual:2003tq}
Bruzual G.,  Charlot S.,  2003, \mn@doi [Monthly Notices of the Royal Astronomical Society] {10.1046/j.1365-8711.2003.06897.x}, 344, 1000

\bibitem[\protect\citeauthoryear{{Cabanac} et~al.,}{{Cabanac} et~al.}{2007}]{Cabanac2007}
{Cabanac} R.~A.,  et~al., 2007, \mn@doi [Astronomy \& Astrophysics] {10.1051/0004-6361:20065810}, \href {https://ui.adsabs.harvard.edu/abs/2007A&A...461..813C} {461, 813}

\bibitem[\protect\citeauthoryear{Cao et~al.}{Cao et~al.}{2024}]{Cao:2023bnl}
Cao X.,  et~al., 2024, \mn@doi [Monthly Notices of the Royal Astronomical Society] {10.1093/mnras/stae1865}, 533, 1960

\bibitem[\protect\citeauthoryear{Casey et~al.}{Casey et~al.}{2023}]{Casey:2022amu}
Casey C.~M.,  et~al., 2023, \mn@doi [The Astrophysical Journal] {10.3847/1538-4357/acc2bc}, 954, 31

\bibitem[\protect\citeauthoryear{{Casey} et~al.,}{{Casey} et~al.}{2024}]{Casey2024}
{Casey} C.~M.,  et~al., 2024, \mn@doi [The Astrophysical Journal] {10.3847/1538-4357/ad2075}, \href {https://ui.adsabs.harvard.edu/abs/2024ApJ...965...98C} {965, 98}

\bibitem[\protect\citeauthoryear{{Chevallard} \& {Charlot}}{{Chevallard} \& {Charlot}}{2016}]{Chevallard2016}
{Chevallard} J.,  {Charlot} S.,  2016, \mn@doi [Monthly Notices of the Royal Astronomical Society] {10.1093/mnras/stw1756}, \href {https://ui.adsabs.harvard.edu/abs/2016MNRAS.462.1415C} {462, 1415}

\bibitem[\protect\citeauthoryear{{Choi}, {Park}  \& {Vogeley}}{{Choi} et~al.}{2007}]{Choi2007}
{Choi} Y.-Y.,  {Park} C.,   {Vogeley} M.~S.,  2007, \mn@doi [The Astrophysical Journal] {10.1086/511060}, \href {https://ui.adsabs.harvard.edu/abs/2007ApJ...658..884C} {658, 884}

\bibitem[\protect\citeauthoryear{Collett}{Collett}{2015}]{Collett:2015roa}
Collett T.~E.,  2015, \mn@doi [The Astrophysical Journal] {10.1088/0004-637X/811/1/20}, 811, 20

\bibitem[\protect\citeauthoryear{{Connolly} et~al.,}{{Connolly} et~al.}{2010}]{Connolly2010}
{Connolly} A.~J.,  et~al., 2010, in {Angeli} G.~Z.,  {Dierickx} P.,  eds,  Society of Photo-Optical Instrumentation Engineers (SPIE) Conference Series Vol. 7738, Modeling, Systems Engineering, and Project Management for Astronomy IV. p. 77381O, \mn@doi{10.1117/12.857819}

\bibitem[\protect\citeauthoryear{De~Lucia, Springel, White, Croton  \& Kauffmann}{De~Lucia et~al.}{2006}]{DeLucia:2005yhi}
De~Lucia G.,  Springel V.,  White S. D.~M.,  Croton D.,   Kauffmann G.,  2006, \mn@doi [Monthly Notices of the Royal Astronomical Society] {10.1111/j.1365-2966.2005.09879.x}, 366, 499

\bibitem[\protect\citeauthoryear{{Djorgovski} \& {Davis}}{{Djorgovski} \& {Davis}}{1987}]{Djorgovski1987}
{Djorgovski} S.,  {Davis} M.,  1987, \mn@doi [The Astrophysical Journal] {10.1086/164948}, \href {https://ui.adsabs.harvard.edu/abs/1987ApJ...313...59D} {313, 59}

\bibitem[\protect\citeauthoryear{Drakos et~al.}{Drakos et~al.}{2022}]{Drakos:2021bsb}
Drakos N.~E.,  et~al., 2022, \mn@doi [The Astrophysical Journal] {10.3847/1538-4357/ac46fb}, 926, 194

\bibitem[\protect\citeauthoryear{{Dressler}, {Lynden-Bell}, {Burstein}, {Davies}, {Faber}, {Terlevich}  \& {Wegner}}{{Dressler} et~al.}{1987}]{Dressler1987}
{Dressler} A.,  {Lynden-Bell} D.,  {Burstein} D.,  {Davies} R.~L.,  {Faber} S.~M.,  {Terlevich} R.,   {Wegner} G.,  1987, \mn@doi [The Astrophysical Journal] {10.1086/164947}, \href {https://ui.adsabs.harvard.edu/abs/1987ApJ...313...42D} {313, 42}

\bibitem[\protect\citeauthoryear{Duboscq, Hogg, Fleury  \& Larena}{Duboscq et~al.}{2024}]{Duboscq:2024asf}
Duboscq T.,  Hogg N.~B.,  Fleury P.,   Larena J.,  2024, \mn@doi [JCAP] {10.1088/1475-7516/2024/08/021}, 08, 021

\bibitem[\protect\citeauthoryear{{Eisenstein} et~al.}{{Eisenstein} et~al.}{2023}]{Eisenstein2023}
{Eisenstein} D.~J.,  et~al., 2023, {Overview of the JWST Advanced Deep Extragalactic Survey (JADES)} (\mn@eprint {arXiv} {2306.02465})

\bibitem[\protect\citeauthoryear{Faure et~al.}{Faure et~al.}{2008}]{Faure:2008dt}
Faure C.,  et~al., 2008, \mn@doi [The Astrophysical Journal Supplement] {10.1086/526426}, 176, 19

\bibitem[\protect\citeauthoryear{Ferrami \& Wyithe}{Ferrami \& Wyithe}{2024}]{Ferrami:2024obm}
Ferrami G.,  Wyithe J. S.~B.,  2024, \mn@doi [Monthly Notices of the Royal Astronomical Society] {10.1093/mnras/stae1607}, 532, 1832

\bibitem[\protect\citeauthoryear{Fleury, Larena  \& Uzan}{Fleury et~al.}{2021}]{Fleury:2021tke}
Fleury P.,  Larena J.,   Uzan J.-P.,  2021, \mn@doi [Journal of Cosmology and Astroparticle Physics] {10.1088/1475-7516/2021/08/024}, 08, 024

\bibitem[\protect\citeauthoryear{{Franco} et~al.,}{{Franco} et~al.}{2024}]{Franco2024}
{Franco} M.,  et~al., 2024, \mn@doi [The Astrophysical Journal] {10.3847/1538-4357/ad5e6a}, \href {https://ui.adsabs.harvard.edu/abs/2024ApJ...973...23F} {973, 23}

\bibitem[\protect\citeauthoryear{Garvin, Kruk, Cornen, Bhatawdekar, Ca\~nameras  \& Mer\'\i{}n}{Garvin et~al.}{2022}]{Garvin:2022gaq}
Garvin E.~O.,  Kruk S.,  Cornen C.,  Bhatawdekar R.,  Ca\~nameras R.,   Mer\'\i{}n B.,  2022, \mn@doi [Astronomy \& Astrophysics] {10.1051/0004-6361/202243745}, 667, A141

\bibitem[\protect\citeauthoryear{Gavazzi, Marshall, Treu  \& Sonnenfeld}{Gavazzi et~al.}{2014}]{Gavazzi:2014nza}
Gavazzi R.,  Marshall P.~J.,  Treu T.,   Sonnenfeld A.,  2014, \mn@doi [The Astrophysical Journal] {10.1088/0004-637X/785/2/144}, 785, 144

\bibitem[\protect\citeauthoryear{{Gelli}, {Salvadori}, {Ferrara}, {Pallottini}  \& {Carniani}}{{Gelli} et~al.}{2023}]{Gelli2023}
{Gelli} V.,  {Salvadori} S.,  {Ferrara} A.,  {Pallottini} A.,   {Carniani} S.,  2023, \mn@doi [The Astrophysical Journal: Letters] {10.3847/2041-8213/acee80}, \href {https://ui.adsabs.harvard.edu/abs/2023ApJ...954L..11G} {954, L11}

\bibitem[\protect\citeauthoryear{Hartigan \& Hartigan}{Hartigan \& Hartigan}{1985}]{Hartigan1985}
Hartigan J.~A.,  Hartigan P.~M.,  1985, Annals of Statistics, 13, 70

\bibitem[\protect\citeauthoryear{He et~al.}{He et~al.}{2024}]{He:2024udi}
He Q.,  et~al., 2024, \mn@doi [Monthly Notices of the Royal Astronomical Society] {10.1093/mnras/stae1577}, 532, 2441

\bibitem[\protect\citeauthoryear{Heymans et~al.}{Heymans et~al.}{2021}]{Heymans:2020gsg}
Heymans C.,  et~al., 2021, \mn@doi [Astronomy \& Astrophysics] {10.1051/0004-6361/202039063}, 646, A140

\bibitem[\protect\citeauthoryear{{Hezaveh} et~al.,}{{Hezaveh} et~al.}{2013}]{Hezaveh2013}
{Hezaveh} Y.~D.,  et~al., 2013, \mn@doi [The Astrophysical Journal] {10.1088/0004-637X/767/2/132}, \href {https://ui.adsabs.harvard.edu/abs/2013ApJ...767..132H} {767, 132}

\bibitem[\protect\citeauthoryear{Hogg}{Hogg}{2024}]{Hogg:2023khs}
Hogg N.~B.,  2024, \mn@doi [Monthly Notices of the Royal Astronomical Society: Letters] {10.1093/mnrasl/slae005}, 528, L95

\bibitem[\protect\citeauthoryear{Hogg, Fleury, Larena  \& Martinelli}{Hogg et~al.}{2023}]{Hogg:2022ycw}
Hogg N.~B.,  Fleury P.,  Larena J.,   Martinelli M.,  2023, \mn@doi [Monthly Notices of the Royal Astronomical Society] {10.1093/mnras/stad512}, 520, 5982

\bibitem[\protect\citeauthoryear{Hogg, Shajib, Johnson  \& Larena}{Hogg et~al.}{2025}]{Hogg:2025wac}
Hogg N.~B.,  Shajib A.~J.,  Johnson D.,   Larena J.,  2025, {Line-of-sight shear in SLACS strong lenses} (\mn@eprint {arXiv} {2501.16292})

\bibitem[\protect\citeauthoryear{Holloway, Verma, Marshall, More  \& Tecza}{Holloway et~al.}{2023}]{Holloway:2023axl}
Holloway P.,  Verma A.,  Marshall P.~J.,  More A.,   Tecza M.,  2023, \mn@doi [Monthly Notices of the Royal Astronomical Society] {10.1093/mnras/stad2371}, 525, 2341

\bibitem[\protect\citeauthoryear{{Holloway} et~al.,}{{Holloway} et~al.}{2025}]{EuclidQ1E}
{Holloway} P.,  et~al., 2025, \mn@doi [arXiv e-prints] {10.48550/arXiv.2503.15328}, \href {https://ui.adsabs.harvard.edu/abs/2025arXiv250315328E} {p. arXiv:2503.15328}

\bibitem[\protect\citeauthoryear{Huterer, Keeton  \& Ma}{Huterer et~al.}{2005}]{Huterer:2004jh}
Huterer D.,  Keeton C.~R.,   Ma C.-P.,  2005, \mn@doi [The Astrophysical Journal] {10.1086/429153}, 624, 34

\bibitem[\protect\citeauthoryear{Hyde \& Bernardi}{Hyde \& Bernardi}{2009a}]{Hyde:2008yf}
Hyde J.~B.,  Bernardi M.,  2009a, \mn@doi [Monthly Notices of the Royal Astronomical Society] {10.1111/j.1365-2966.2009.14445.x}, 394, 1978

\bibitem[\protect\citeauthoryear{Hyde \& Bernardi}{Hyde \& Bernardi}{2009b}]{Hyde:2008yh}
Hyde J.~B.,  Bernardi M.,  2009b, \mn@doi [Monthly Notices of the Royal Astronomical Society] {10.1111/j.1365-2966.2009.14783.x}, 396, 1171

\bibitem[\protect\citeauthoryear{Jackson}{Jackson}{2008}]{Jackson:2008my}
Jackson N.,  2008, \mn@doi [Monthly Notices of the Royal Astronomical Society] {10.1111/j.1365-2966.2008.13629.x}, 389, 1311

\bibitem[\protect\citeauthoryear{{Jin} et~al.,}{{Jin} et~al.}{2018}]{Jin2018}
{Jin} S.,  et~al., 2018, \mn@doi [The Astrophysical Journal] {10.3847/1538-4357/aad4af}, \href {https://ui.adsabs.harvard.edu/abs/2018ApJ...864...56J} {864, 56}

\bibitem[\protect\citeauthoryear{{Jin} et~al.,}{{Jin} et~al.}{2024}]{Jin2024}
{Jin} S.,  et~al., 2024, \mn@doi [Astronomy \& Astrophysics] {10.1051/0004-6361/202451445}, \href {https://ui.adsabs.harvard.edu/abs/2024A&A...690L..16J} {690, L16}

\bibitem[\protect\citeauthoryear{Kochanek}{Kochanek}{2004}]{Kochanek:2004ua}
Kochanek C.~S.,  2004, in {33rd Advanced Saas Fee Course on Gravitational Lensing: Strong, Weak, and Micro}.  (\mn@eprint {arXiv} {astro-ph/0407232})

\bibitem[\protect\citeauthoryear{{Koekemoer} et~al.,}{{Koekemoer} et~al.}{2007}]{Koekemoer2007}
{Koekemoer} A.~M.,  et~al., 2007, \mn@doi [The Astrophysical Journal Supplement] {10.1086/520086}, \href {https://ui.adsabs.harvard.edu/abs/2007ApJS..172..196K} {172, 196}

\bibitem[\protect\citeauthoryear{{Kormann}, {Schneider}  \& {Bartelmann}}{{Kormann} et~al.}{1994}]{Kormann1994}
{Kormann} R.,  {Schneider} P.,   {Bartelmann} M.,  1994, Astronomy \& Astrophysics, \href {https://ui.adsabs.harvard.edu/abs/1994A&A...284..285K} {284, 285}

\bibitem[\protect\citeauthoryear{Lanusse, Ma, Li, Collett, Li, Ravanbakhsh, Mandelbaum  \& Poczos}{Lanusse et~al.}{2018}]{Lanusse:2017vha}
Lanusse F.,  Ma Q.,  Li N.,  Collett T.~E.,  Li C.-L.,  Ravanbakhsh S.,  Mandelbaum R.,   Poczos B.,  2018, \mn@doi [Monthly Notices of the Royal Astronomical Society] {10.1093/mnras/stx1665}, 473, 3895

\bibitem[\protect\citeauthoryear{Lee \& Kim}{Lee \& Kim}{2014}]{Lee:2014yxa}
Lee D.-W.,  Kim S.-J.,  2014, \mn@doi [Monthly Notices of the Royal Astronomical Society] {10.1093/mnras/stu970}, 443, 328

\bibitem[\protect\citeauthoryear{{Li}, {Collett}, {Krawczyk}  \& {Enzi}}{{Li} et~al.}{2024}]{Li2024}
{Li} T.,  {Collett} T.~E.,  {Krawczyk} C.~M.,   {Enzi} W.,  2024, \mn@doi [Monthly Notices of the Royal Astronomical Society] {10.1093/mnras/stad3514}, \href {https://ui.adsabs.harvard.edu/abs/2024MNRAS.527.5311L} {527, 5311}

\bibitem[\protect\citeauthoryear{{Li} et~al.,}{{Li} et~al.}{2025}]{EuclidQ1D}
{Li} T.,  et~al., 2025, \mn@doi [arXiv e-prints] {10.48550/arXiv.2503.15327}, \href {https://ui.adsabs.harvard.edu/abs/2025arXiv250315327E} {p. arXiv:2503.15327}

\bibitem[\protect\citeauthoryear{{Lin} et~al.,}{{Lin} et~al.}{2023}]{Lin2023}
{Lin} X.,  et~al., 2023, \mn@doi [The Astrophysical Journal: Letters] {10.3847/2041-8213/aca1c4}, \href {https://ui.adsabs.harvard.edu/abs/2023ApJ...944L..59L} {944, L59}

\bibitem[\protect\citeauthoryear{{Lines} et~al.,}{{Lines} et~al.}{2025}]{EuclidQ1C}
{Lines} N.~E.~P.,  et~al., 2025, \mn@doi [arXiv e-prints] {10.48550/arXiv.2503.15326}, \href {https://ui.adsabs.harvard.edu/abs/2025arXiv250315326E} {p. arXiv:2503.15326}

\bibitem[\protect\citeauthoryear{Mahler et~al.}{Mahler et~al.}{2025}]{Mahler2025}
Mahler G.,  et~al., 2025, {The COSMOS-Web Lens Survey (COWLS) II: depth, resolution, and NIR coverage from JWST reveal 17 spectacular lenses} (\mn@eprint {arXiv} {2503.08782})

\bibitem[\protect\citeauthoryear{{Marshall}, {Hogg}, {Moustakas}, {Fassnacht}, {Brada{\v{c}}}, {Schrabback}  \& {Blandford}}{{Marshall} et~al.}{2009}]{Marshall2009}
{Marshall} P.~J.,  {Hogg} D.~W.,  {Moustakas} L.~A.,  {Fassnacht} C.~D.,  {Brada{\v{c}}} M.,  {Schrabback} T.,   {Blandford} R.~D.,  2009, \mn@doi [The Astrophysical Journal] {10.1088/0004-637X/694/2/924}, \href {https://ui.adsabs.harvard.edu/abs/2009ApJ...694..924M} {694, 924}

\bibitem[\protect\citeauthoryear{Marshall et~al.}{Marshall et~al.}{2016}]{Marshall:2015fsa}
Marshall P.~J.,  et~al., 2016, \mn@doi [Monthly Notices of the Royal Astronomical Society] {10.1093/mnras/stv2009}, 455, 1171

\bibitem[\protect\citeauthoryear{Mason et~al.,}{Mason et~al.}{2015}]{Mason:2015wla}
Mason C.~A.,  et~al., 2015, \mn@doi [The Astrophysical Journal] {10.1088/0004-637X/805/1/79}, 805, 79

\bibitem[\protect\citeauthoryear{Mellier et~al.}{Mellier et~al.}{2025}]{Mellier2024}
Mellier Y.,  et~al., 2025, \mn@doi [Astronomy \& Astrophysics] {10.1051/0004-6361/202450810}, 697, A1

\bibitem[\protect\citeauthoryear{{Mercier} et~al.,}{{Mercier} et~al.}{2024}]{Mercier2024}
{Mercier} W.,  et~al., 2024, \mn@doi [Astronomy \& Astrophysics] {10.1051/0004-6361/202348095}, \href {https://ui.adsabs.harvard.edu/abs/2024A&A...687A..61M} {687, A61}

\bibitem[\protect\citeauthoryear{{More}, {Cabanac}, {More}, {Alard}, {Limousin}, {Kneib}, {Gavazzi}  \& {Motta}}{{More} et~al.}{2012}]{More2012}
{More} A.,  {Cabanac} R.,  {More} S.,  {Alard} C.,  {Limousin} M.,  {Kneib} J.~P.,  {Gavazzi} R.,   {Motta} V.,  2012, \mn@doi [The Astrophysical Journal] {10.1088/0004-637X/749/1/38}, \href {https://ui.adsabs.harvard.edu/abs/2012ApJ...749...38M} {749, 38}

\bibitem[\protect\citeauthoryear{More et~al.}{More et~al.}{2016}]{More:2015sfa}
More A.,  et~al., 2016, \mn@doi [Monthly Notices of the Royal Astronomical Society] {10.1093/mnras/stv1965}, 455, 1191

\bibitem[\protect\citeauthoryear{More et~al.,}{More et~al.}{2024}]{More:2024uvq}
More A.,  et~al., 2024, \mn@doi [Monthly Notices of the Royal Astronomical Society] {10.1093/mnras/stae1597}, 533, 525

\bibitem[\protect\citeauthoryear{{Nagam} et~al.}{{Nagam} et~al.}{2025}]{Nagam2025}
{Nagam} B.~C.,  et~al., 2025, {Euclid: Finding strong gravitational lenses in the Early Release Observations using convolutional neural networks} (\mn@eprint {arXiv} {2502.09802}), \mn@doi{10.48550/arXiv.2502.09802}

\bibitem[\protect\citeauthoryear{{Newman} et~al.}{{Newman} et~al.}{2025}]{Newman2025}
{Newman} S.~L.,  et~al., 2025 (\mn@eprint {arXiv} {2501.03133})

\bibitem[\protect\citeauthoryear{Nightingale et~al.,}{Nightingale et~al.}{2021}]{pyautolens}
Nightingale J.~W.,  et~al., 2021, \mn@doi [Journal of Open Source Software] {10.21105/joss.02825}, 6, 2825

\bibitem[\protect\citeauthoryear{Nightingale et~al.}{Nightingale et~al.}{2025}]{Nightingale2025}
Nightingale J.,  et~al., 2025, {The COSMOS-Web Lens Survey (COWLS) I: Discovery of \ensuremath{>}100 high redshift strong lenses in contiguous JWST imaging} (\mn@eprint {arXiv} {2503.08777})

\bibitem[\protect\citeauthoryear{{O'Riordan}}{{O'Riordan}}{2025}]{ORiordan2025}
{O'Riordan} C.~M.,  2025, \mn@doi [Astronomy \& Astrophysics] {10.1051/0004-6361/202453014}, \href {https://ui.adsabs.harvard.edu/abs/2025A&A...694A.145O} {694, A145}

\bibitem[\protect\citeauthoryear{O'Riordan, Despali, Vegetti, Lovell  \& Molin\'e}{O'Riordan et~al.}{2023}]{ORiordan:2022qds}
O'Riordan C.~M.,  Despali G.,  Vegetti S.,  Lovell M.~R.,   Molin\'e A.,  2023, \mn@doi [Monthly Notices of the Royal Astronomical Society] {10.1093/mnras/stad650}, 521, 2342

\bibitem[\protect\citeauthoryear{Pearce-Casey et~al.}{Pearce-Casey et~al.}{2024}]{Euclid:2024juc}
Pearce-Casey R.,  et~al., 2024, {Euclid: Searches for strong gravitational lenses using convolutional neural nets in Early Release Observations of the Perseus field} (\mn@eprint {arXiv} {2411.16808})

\bibitem[\protect\citeauthoryear{Pearson et~al.}{Pearson et~al.}{2023}]{Pearson:2023kfx}
Pearson J.,  et~al., 2023, \mn@doi [Monthly Notices of the Royal Astronomical Society] {10.1093/mnras/stad3916}, 527, 12044

\bibitem[\protect\citeauthoryear{{Pourrahmani}, {Nayyeri}  \& {Cooray}}{{Pourrahmani} et~al.}{2018}]{Pourrahmani2018}
{Pourrahmani} M.,  {Nayyeri} H.,   {Cooray} A.,  2018, \mn@doi [The Astrophysical Journal] {10.3847/1538-4357/aaae6a}, \href {https://ui.adsabs.harvard.edu/abs/2018ApJ...856...68P} {856, 68}

\bibitem[\protect\citeauthoryear{Razali \& Wah}{Razali \& Wah}{2011}]{Razali2011}
Razali N.~M.,  Wah Y.~B.,  2011, Journal of Statistical Modelling and Analytics, 2, 21

\bibitem[\protect\citeauthoryear{Refsdal}{Refsdal}{1964}]{Refsdal1964b}
Refsdal S.,  1964, \mn@doi [Monthly Notices of the Royal Astronomical Society] {10.1093/mnras/128.4.307}, 128, 307

\bibitem[\protect\citeauthoryear{Rodrigo, Solano  \& Bayo}{Rodrigo et~al.}{2012}]{Rodrigo2012}
Rodrigo C.,  Solano E.,   Bayo A.,  2012, IVOA Working Draft

\bibitem[\protect\citeauthoryear{Rojas et~al.}{Rojas et~al.}{2023}]{DES:2023hxm}
Rojas K.,  et~al., 2023, \mn@doi [Monthly Notices of the Royal Astronomical Society] {10.1093/mnras/stad1680}, 523, 4413

\bibitem[\protect\citeauthoryear{{Rojas} et~al.,}{{Rojas} et~al.}{2025}]{EuclidQ1B}
{Rojas} K.,  et~al., 2025, \mn@doi [arXiv e-prints] {10.48550/arXiv.2503.15325}, \href {https://ui.adsabs.harvard.edu/abs/2025arXiv250315325E} {p. arXiv:2503.15325}

\bibitem[\protect\citeauthoryear{{Sainz de Murieta}, {Collett}, {Magee}, {Pierel}, {Enzi}, {Lokken}, {Gagliano}  \& {Ryczanowski}}{{Sainz de Murieta} et~al.}{2024}]{Sainz2024}
{Sainz de Murieta} A.,  {Collett} T.~E.,  {Magee} M.~R.,  {Pierel} J. D.~R.,  {Enzi} W. J.~R.,  {Lokken} M.,  {Gagliano} A.,   {Ryczanowski} D.,  2024, \mn@doi [Monthly Notices of the Royal Astronomical Society] {10.1093/mnras/stae2486}, \href {https://ui.adsabs.harvard.edu/abs/2024MNRAS.535.2523S} {535, 2523}

\bibitem[\protect\citeauthoryear{Scoville et~al.}{Scoville et~al.}{2007}]{Scoville:2006vq}
Scoville N.,  et~al., 2007, \mn@doi [The Astrophysical Journal Supplement] {10.1086/516585}, 172, 1

\bibitem[\protect\citeauthoryear{{Shajib}, {Treu}, {Birrer}  \& {Sonnenfeld}}{{Shajib} et~al.}{2021}]{Shajib2021}
{Shajib} A.~J.,  {Treu} T.,  {Birrer} S.,   {Sonnenfeld} A.,  2021, \mn@doi [Monthly Notices of the Royal Astronomical Society] {10.1093/mnras/stab536}, \href {https://ui.adsabs.harvard.edu/abs/2021MNRAS.503.2380S} {503, 2380}

\bibitem[\protect\citeauthoryear{{Shuntov} et~al.,}{{Shuntov} et~al.}{2025}]{Shuntov2025}
{Shuntov} M.,  et~al., 2025, \mn@doi [arXiv e-prints] {10.48550/arXiv.2502.20136}, \href {https://ui.adsabs.harvard.edu/abs/2025arXiv250220136S} {p. arXiv:2502.20136}

\bibitem[\protect\citeauthoryear{Silverman}{Silverman}{1981}]{Silverman1981}
Silverman B.~W.,  1981, Journal of the Royal Statistical Society, 43, 97

\bibitem[\protect\citeauthoryear{{Sonnenfeld}}{{Sonnenfeld}}{2022}]{Sonnenfeld2022}
{Sonnenfeld} A.,  2022, \mn@doi [Astronomy \& Astrophysics] {10.1051/0004-6361/202142301}, \href {https://ui.adsabs.harvard.edu/abs/2022A&A...659A.132S} {659, A132}

\bibitem[\protect\citeauthoryear{{Sonnenfeld}}{{Sonnenfeld}}{2024}]{Sonnenfeld2024}
{Sonnenfeld} A.,  2024, \mn@doi [Astronomy \& Astrophysics] {10.1051/0004-6361/202451341}, \href {https://ui.adsabs.harvard.edu/abs/2024A&A...690A.325S} {690, A325}

\bibitem[\protect\citeauthoryear{{Sonnenfeld} \& {Cautun}}{{Sonnenfeld} \& {Cautun}}{2021}]{Sonnenfeld2021b}
{Sonnenfeld} A.,  {Cautun} M.,  2021, \mn@doi [Astronomy \& Astrophysics] {10.1051/0004-6361/202140549}, \href {https://ui.adsabs.harvard.edu/abs/2021A&A...651A..18S} {651, A18}

\bibitem[\protect\citeauthoryear{{Sonnenfeld} et~al.,}{{Sonnenfeld} et~al.}{2020}]{Sonnenfeld2020}
{Sonnenfeld} A.,  et~al., 2020, \mn@doi [Astronomy \& Astrophysics] {10.1051/0004-6361/202038067}, \href {https://ui.adsabs.harvard.edu/abs/2020A&A...642A.148S} {642, A148}

\bibitem[\protect\citeauthoryear{{Sonnenfeld}, {Li}, {Despali}, {Gavazzi}, {Shajib}  \& {Taylor}}{{Sonnenfeld} et~al.}{2023}]{Sonnenfeld2023b}
{Sonnenfeld} A.,  {Li} S.-S.,  {Despali} G.,  {Gavazzi} R.,  {Shajib} A.~J.,   {Taylor} E.~N.,  2023, \mn@doi [Astronomy \& Astrophysics] {10.1051/0004-6361/202346026}, \href {https://ui.adsabs.harvard.edu/abs/2023A&A...678A...4S} {678, A4}

\bibitem[\protect\citeauthoryear{Sprent}{Sprent}{1998}]{Sprent1998}
Sprent P.,  1998, {Data Driven Statistical Methods}.
Chapman \& Hall

\bibitem[\protect\citeauthoryear{Suyu et~al.}{Suyu et~al.}{2014}]{Suyu:2013kha}
Suyu S.~H.,  et~al., 2014, \mn@doi [The Astrophysical Journal: Letters] {10.1088/2041-8205/788/2/L35}, 788, L35

\bibitem[\protect\citeauthoryear{{Swinbank} et~al.,}{{Swinbank} et~al.}{2015}]{Swinbank2015}
{Swinbank} A.~M.,  et~al., 2015, \mn@doi [The Astrophysical Journal: Letters] {10.1088/2041-8205/806/1/L17}, \href {https://ui.adsabs.harvard.edu/abs/2015ApJ...806L..17S} {806, L17}

\bibitem[\protect\citeauthoryear{{Tan} et~al.,}{{Tan} et~al.}{2024}]{Tan2024}
{Tan} C.~Y.,  et~al., 2024, \mn@doi [Monthly Notices of the Royal Astronomical Society] {10.1093/mnras/stae884}, \href {https://ui.adsabs.harvard.edu/abs/2024MNRAS.530.1474T} {530, 1474}

\bibitem[\protect\citeauthoryear{Vegetti, Lagattuta, McKean, Auger, Fassnacht  \& Koopmans}{Vegetti et~al.}{2012}]{Vegetti2012}
Vegetti S.,  Lagattuta D.~J.,  McKean J.~P.,  Auger M.~W.,  Fassnacht C.~D.,   Koopmans L. V.~E.,  2012, \mn@doi [Nature] {10.1038/nature10669}, 481, 341

\bibitem[\protect\citeauthoryear{{Walmsley} et~al.,}{{Walmsley} et~al.}{2025}]{EuclidQ1A}
{Walmsley} M.,  et~al., 2025, \mn@doi [arXiv e-prints] {10.48550/arXiv.2503.15324}, \href {https://ui.adsabs.harvard.edu/abs/2025arXiv250315324E} {p. arXiv:2503.15324}

\bibitem[\protect\citeauthoryear{{Weiner}, {Serjeant}  \& {Sedgwick}}{{Weiner} et~al.}{2020}]{Weiner2020}
{Weiner} C.,  {Serjeant} S.,   {Sedgwick} C.,  2020, \mn@doi [Research Notes of the American Astronomical Society] {10.3847/2515-5172/abc4ea}, \href {https://ui.adsabs.harvard.edu/abs/2020RNAAS...4..190W} {4, 190}

\bibitem[\protect\citeauthoryear{{Williams} et~al.,}{{Williams} et~al.}{1996}]{williams1996}
{Williams} R.~E.,  et~al., 1996, \mn@doi [The Astronomical Journal] {10.1086/118105}, \href {https://ui.adsabs.harvard.edu/abs/1996AJ....112.1335W} {112, 1335}

\bibitem[\protect\citeauthoryear{{Williams} et~al.,}{{Williams} et~al.}{2018}]{Williams2018}
{Williams} C.~C.,  et~al., 2018, \mn@doi [The Astrophysical Journal Supplement] {10.3847/1538-4365/aabcbb}, \href {https://ui.adsabs.harvard.edu/abs/2018ApJS..236...33W} {236, 33}

\bibitem[\protect\citeauthoryear{Wong et~al.}{Wong et~al.}{2020}]{Wong:2019kwg}
Wong K.~C.,  et~al., 2020, \mn@doi [Monthly Notices of the Royal Astronomical Society] {10.1093/mnras/stz3094}, 498

\bibitem[\protect\citeauthoryear{York et~al.}{York et~al.}{2000}]{SDSS:2000hjo}
York D.~G.,  et~al., 2000, \mn@doi [The Astronomical Journal] {10.1086/301513}, 120, 1579

\bibitem[\protect\citeauthoryear{{de Vaucouleurs}}{{de Vaucouleurs}}{1948}]{deVaucouleurs1948}
{de Vaucouleurs} G.,  1948, Annales d'Astrophysique, \href {https://ui.adsabs.harvard.edu/abs/1948AnAp...11..247D} {11, 247}

\makeatother
\end{thebibliography}

\noindent\rule{\columnwidth}{0.4pt}
% List of institutions
$^{1}$\LUPM\\
$^{2}$\Newcastle\\
$^{3}$\DurhamICC\\
$^{4}$\Northeastern\\
$^{5}$\Liege\\
$^{6}$\DurhamCEA\\
$^{7}$\Aalto\\
$^{8}$\Helsinki\\
$^{9}$\JPL\\
$^{10}$\UTAustin\\
$^{11}$\PMO\\
$^{12}$\DAWN\\
$^{13}$\NBI\\
$^{14}$\IAP\\
$^{15}$\LAM\\
$^{16}$\UCSB\\
$^{17}$\Rochester\\
$^{18}$\STScI\\
$^{19}$\UCSC\\
$^{20}$\Hawaii\\
$^{21}$\Caltech\\
$^{22}$\UCR\\
$^{23}$\ICG

\appendix
\section{Summary of statistical tests} \label{sec:stattables}
In this Appendix we tabulate the results of the statistical tests used to compare the forecast and observed data. For each test, the null hypothesis is that the forecast and observed data were drawn from the same underlying distribution. \autoref{tab:specstats} shows the test results for the comparison between the forecast and the \spectacularlenses spectacular lenses; \autoref{tab:ntotstats} shows the test results for the comparison between the forecast and the \Ntotal high confidence lenses; \autoref{tab:ntotweakstats} shows the test results for the comparison between the forecast and the \Ntotalweak high and medium confidence lenses. For the KS test, the null hypothesis is rejected at 95\% confidence if the p-value is less than 0.05; for the AD test, the null hypothesis is rejected if the test statistic is greater than the critical value, which for 95\% confidence is $1.961$. We note that the \texttt{scipy.stats.anderson\_ksamp} method we use to compute the AD test in Python returns p-values of $1.0\times10^{-3}$ as a proxy for very small p-values that cannot be computed exactly.

% note to self: I've computed all these on the full forecast catalogue rather than 107 selected; this inflates the numbers a bit but not significantly

\setlength{\tabcolsep}{0.4em}
\begin{table}
    \centering
    \begin{tabular}{Slcccccl}
             & \multicolumn{3}{c}{KS} &  \multicolumn{3}{c}{AD} \\
    Quantity & Statistic & p-value & H$_0$ & Statistic & p-value & H$_0$ \\
    \hline
    \hline
    $\theta_{\rm E}$             & $0.398$ & $6.1\times 10^{-3}$ &\faClose  & $8.790$   & $1.0\times 10^{-3}$ & \faClose\\
    $z_{\rm lens}$               & $0.218$ & $3.8\times 10^{-1}$ &\faCheck & $0.190$  & $2.5\times 10^{-1}$ & \faCheck\\
    $M_{\rm lens}^{\rm F115W}$   & $0.361$ & $2.2\times 10^{-2}$ &\faClose & $2.864$   & $2.2\times 10^{-2}$ & \faClose\\
    $M_{\rm lens}^{\rm F150W}$   & $0.418$ & $3.3\times 10^{-3}$ &\faClose & $4.229$   & $6.6\times 10^{-3}$ & \faClose\\
    $M_{\rm lens}^{\rm F277W}$   & $0.658$ & $9.5\times 10^{-8}$ &\faClose & $33.390$  & $1.0\times 10^{-3}$ & \faClose\\
    $M_{\rm lens}^{\rm F444W}$   & $0.807$ & $1.7\times 10^{-12}$&\faClose & $48.953$  & $1.0\times 10^{-3}$ & \faClose\\
    $M_{\rm source}^{\rm F115W}$ & $0.328$ & $5.0\times 10^{-2}$ &\faClose & $4.704$  & $4.4\times 10^{-3}$ & \faClose\\
    $M_{\rm source}^{\rm F150W}$ & $0.218$ & $3.5\times 10^{-1}$ &\faCheck & $1.595$  & $7.1\times 10^{-2}$ & \faCheck\\
    $M_{\rm source}^{\rm F277W}$ & $0.304$ & $6.8\times 10^{-2}$  &\faCheck & $2.915$  & $2.1\times 10^{-2}$ & \faClose\\
    $M_{\rm source}^{\rm F444W}$ & $0.195$ & $4.8\times 10^{-1}$ &\faCheck & $1.023$  & $1.2\times 10^{-1}$ & \faCheck\\
    \hline
    \end{tabular}
    \caption{Results of the statistical tests comparing the forecast with the \spectacularlenses spectacular lenses. The `H$_0$' column indicates when the null hypothesis is sustained with a tick (\faCheck) and when it is rejected with a cross (\faClose).}
    \label{tab:specstats}
\end{table}

\begin{table}
    \centering
    \begin{tabular}{Slcccccl}
             & \multicolumn{3}{c}{KS} &  \multicolumn{3}{c}{AD} \\
    Quantity & Statistic & p-value & H$_0$  & Statistic & p-value & H$_0$ \\
    \hline
    \hline
    $\theta_{\rm E}$             & $0.101$ & $5.6 \times 10^{-1}$  &\faCheck & $0.750$  & $1.6\times 10^{-1}$ &\faCheck\\
    $z_{\rm lens}$               & $0.224$ & $5.3 \times 10^{-3}$  &\faClose & $3.479$  & $1.3\times 10^{-2}$ &\faClose\\
    $M_{\rm lens}^{\rm F115W}$   & $0.238$ & $2.8 \times 10^{-3}$  &\faClose & $5.511$  & $2.3\times 10^{-3}$ &\faClose\\
    $M_{\rm lens}^{\rm F150W}$   & $0.268$ & $4.3 \times 10^{-4}$  &\faClose & $5.370$  & $2.6\times10^{-3}$  &\faClose\\
    $M_{\rm lens}^{\rm F277W}$   & $0.517$ & $6.3 \times 10^{-15}$ &\faClose & $69.439$ & $1.0\times 10^{-3}$ &\faClose\\
    $M_{\rm lens}^{\rm F444W}$   & $0.602$ & $1.2 \times 10^{-20}$ &\faClose & $96.381$ & $1.0\times 10^{-3}$ &\faClose\\
    $M_{\rm source}^{\rm F115W}$ & $0.255$ & $1.3 \times 10^{-3}$  &\faClose & $10.647$ & $1.0\times 10^{-3}$ &\faClose\\
    $M_{\rm source}^{\rm F150W}$ & $0.197$ & $2.1 \times 10^{-2}$  &\faClose & $5.966$  & $1.6\times 10^{-3}$ &\faClose\\
    $M_{\rm source}^{\rm F277W}$ & $0.309$ & $2.0 \times 10^{-5}$  &\faClose & $21.328$ & $1.0\times 10^{-3}$ &\faClose\\
    $M_{\rm source}^{\rm F444W}$ & $0.293$ & $6.7 \times 10^{-5}$  &\faClose & $22.360$ & $1.0\times 10^{-3}$ &\faClose\\
    \hline
    \end{tabular}
\caption{Results of the statistical tests comparing the forecast with the \Ntotal high confidence lenses. The `H$_0$' column indicates when the null hypothesis is sustained with a tick (\faCheck) and when it is rejected with a cross (\faClose).}
    \label{tab:ntotstats}
\end{table}

\begin{table}
    \centering
    \begin{tabular}{Slcccccl}
             & \multicolumn{3}{c}{KS} &  \multicolumn{3}{c}{AD} \\
    Quantity & Statistic & p-value & H$_0$  & Statistic & p-value & H$_0$ \\
    \hline
    \hline
    $\theta_{\rm E}$             & $0.061$ & $7.8\times10^{-1}$ &\faCheck & $0.036$   & $2.5\times10^{-1}$ &\faCheck\\
    $z_{\rm lens}$               & $0.171$ & $2.3\times10^{-3}$ &\faClose & $4.714$   & $4.4\times10^{-3}$ &\faClose \\
    $M_{\rm lens}^{\rm F115W}$   & $0.160$ & $7.3\times10^{-3}$ &\faClose & $3.138$   & $1.7\times10^{-2}$ &\faClose \\
    $M_{\rm lens}^{\rm F150W}$   & $0.103$ & $1.8\times10^{-1}$ &\faCheck & $0.832$   & $1.5\times10^{-1}$ &\faCheck\\
    $M_{\rm lens}^{\rm F277W}$   & $0.366$ & $7.1\times10^{-14}$&\faClose & $69.647$  & $1.0\times10^{-3}$ &\faClose\\
    $M_{\rm lens}^{\rm F444W}$   & $0.436$ & $1.0\times10^{-19}$&\faClose & $101.870$ & $1.0\times10^{-3}$ &\faClose\\
    $M_{\rm source}^{\rm F115W}$ & $0.302$ & $9.8\times10^{-9}$ &\faClose & $39.868$  & $1.0\times10^{-3}$ &\faClose\\
    $M_{\rm source}^{\rm F150W}$ & $0.269$ & $2.0\times10^{-7}$ &\faClose & $26.340$  &$1.0\times10^{-3}$  &\faClose\\
    $M_{\rm source}^{\rm F277W}$ & $0.419$ & $3.6\times10^{-18}$&\faClose & $91.412$  &$1.0\times10^{-3}$  &\faClose\\
    $M_{\rm source}^{\rm F444W}$ & $0.410$ & $2.2\times10^{-17}$&\faClose & $92.232$  &$1.0\times10^{-3}$  &\faClose\\
    \hline
    \end{tabular}
\caption{Results of the statistical tests comparing the forecast with the \Ntotalweak high and medium confidence lenses. The `H$_0$' column indicates when the null hypothesis is sustained with a tick (\faCheck) and when it is rejected with a cross (\faClose).}
    \label{tab:ntotweakstats}
\end{table}

\label{lastpage}
\end{document}